\documentclass[12pt]{article}
\input{QCPaul.sty}
\input{MyQcircuit.sty}

\begin{document}

\title{QC Paulinesia}

\author{Robert R. Tucci\\
        P.O. Box 226\\
        Bedford,  MA   01730\\
        tucci@ar-tiste.com}

\date{ \today}

\maketitle

\vskip.5cm
\noindent An archipelago of identities,
formed from the lava of Pauli Matrices,
by the volcanic activity of Quantum Computing.

       \begin{figure}[h]
            \noindent\includegraphics[height=3.7in]{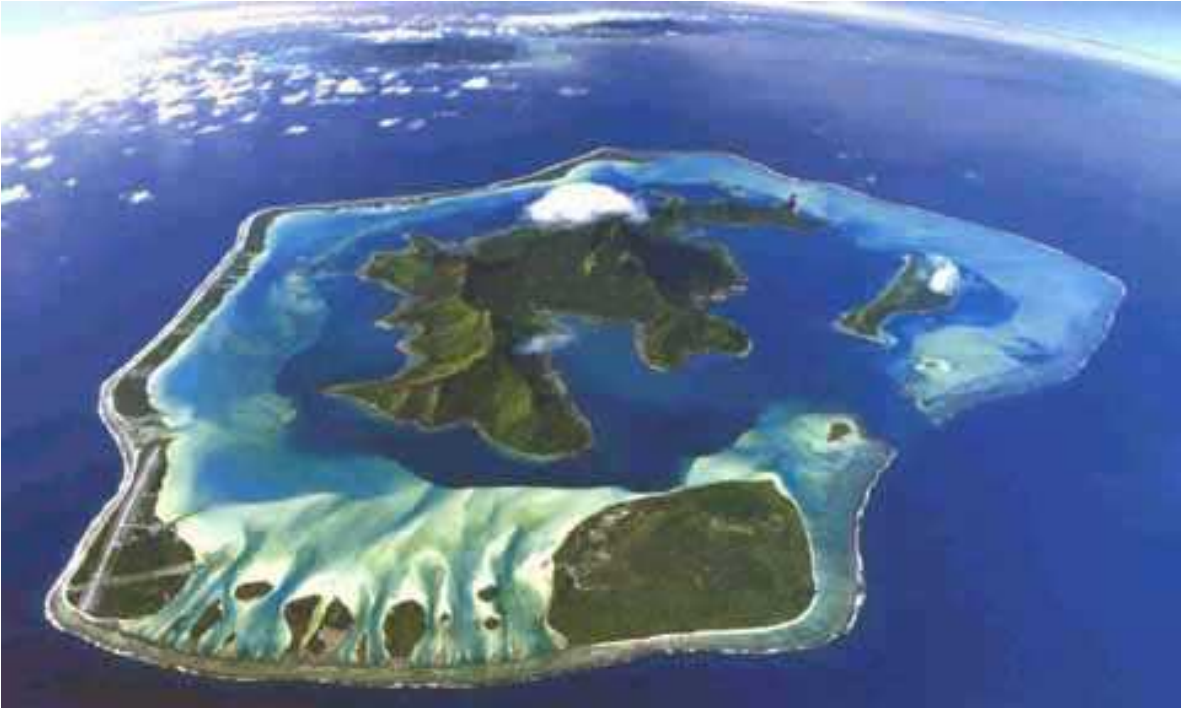}
            \caption{Aerial view of Bora Bora}
      \end{figure}

\noindent\includegraphics[height=8.5in]{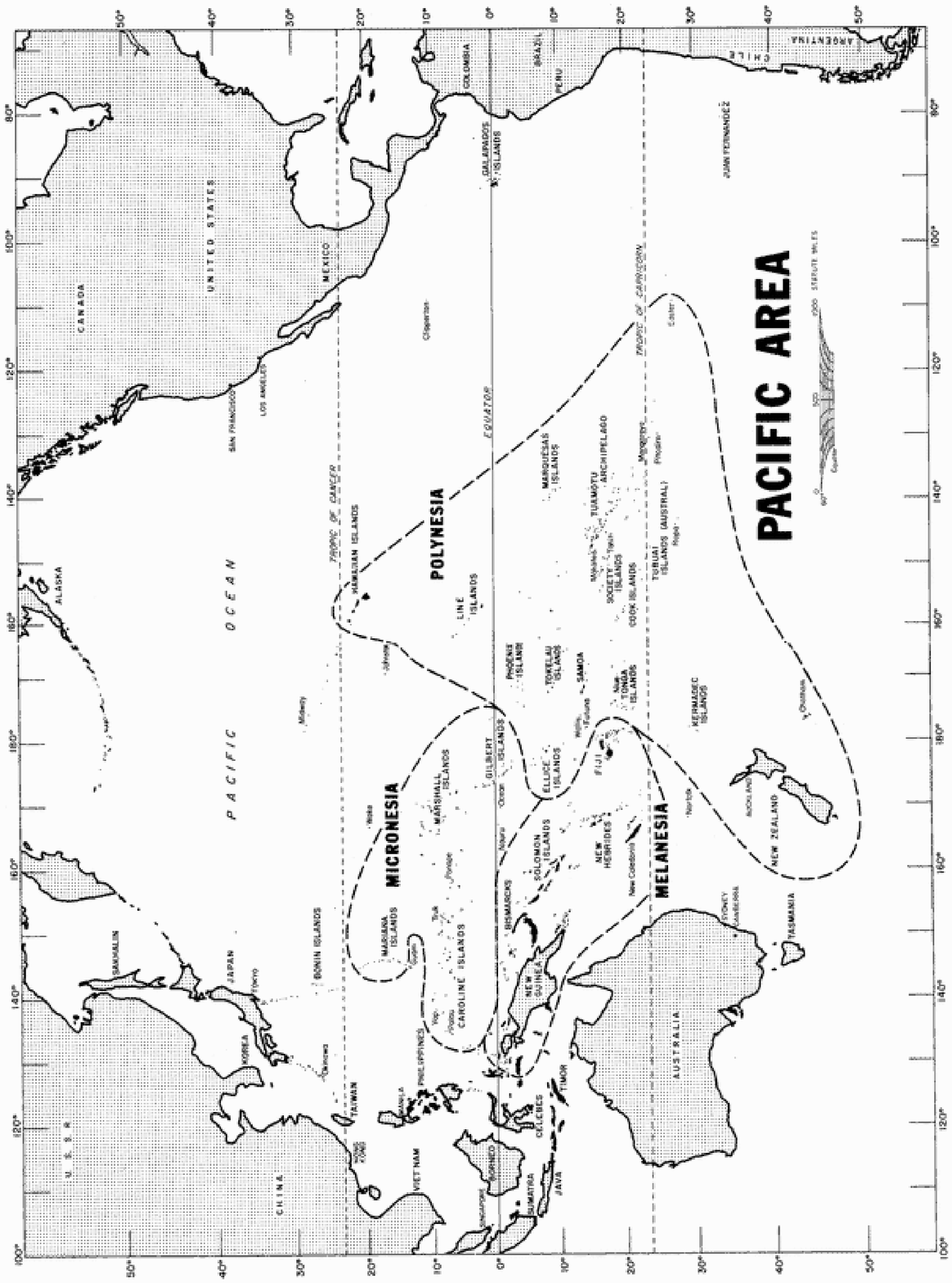}

\tableofcontents

\section{Introduction}

This document is not
a  full course in Quantum Computing.
My goal in producing it
was to create a collection of
qubit circuit identities
that are used
in Quantum Computing.
Mathematicians and
Physicists may consider it
 as being analogous to a
Table of Integrals  or a
 Mathematical Handbook
such as Gradshteyn \& Ryzhik
or Abramowitz \& Stegun.
Computer Programmers may think of it
as a scrapbook of code snippets
that are elegant, instructive,
well documented, and useful.
Electronics experts
may view it as a compendium of
circuits for performing
a large assortment of
tasks.

The vast majority of the  circuit
identities collected in this work
were not discovered for the first time by me, and
I take no credit for discovering them.
In producing this document, I am
acting as a collector,
not as a discoverer.

I plan to
continue adding
qubit circuit identities to this
collection, and to release future versions
of this document
containing the new specimens.
For example, there are some
nice identities involving
quantum error correction and quantum
compiling that I have not included yet,
but which I plan to include in future versions.
Suggestions and comments are
welcomed and appreciated.

This document benefitted greatly
from the wonderful LaTeX
macros: QCircuit (by
B. Eastin, S. T. Flammia)
and XYPic (by K.H. Rose and R.R. Moore),
on which QCircuit is based.

\section{Notation}

Let $Bool = \{0,1\}$.
For integers $a$ and $b$ such that $a\leq b$,
let $Z_{a,b} = \{a, a+1, a+2, \ldots b\}$.

$\delta(x, y)$ and $\delta^x_y$ will both
denote the Kronecker delta function.
It equals one when $x=y$ and zero otherwise.

For any statement ${\cal S}$,
we define the truth function $\theta({\cal S})$
to equal 1 if ${\cal S}$ is true
and 0 if ${\cal S}$ is false.
For example, $\theta(x>0)$ represents the unit step
function and $\delta(x, y)=\theta(x=y)$
the Kronecker delta function.

$\oplus$ will denote
addition mod 2.
Hence, for any $a,b\in Bool$,
$a\oplus b = a + b - 2ab$ and
$(-1)^{a\oplus b} = (-1)^{a+b}$.
When speaking of bits with states
0 and 1, we will often use
an overline to represent
the opposite state:
$\bar{0} = 1$, $\bar{1} = 0$.
Note that if $x, k\in Bool$, then
$\sum_k (-1)^{kx} = 1 + (-1)^x = 2 \delta(x, 0)$.
For $x\in Bool$, $\delta(x,1)=x$.

We will often use $\ns=2^\nb$,
where $\nb$ stands for number of bits and
$\ns$ for number of states.
We will use lower case Latin letters
$a,b,c\ldots \in Bool$ to
represent bit values and
lower case Greek letters
$\alpha, \beta, \gamma, \ldots \in Z_{0, \nb-1} $
to represent bit positions.

Given a binary vector
$\vec{x}\in Bool^\nb$,
if its components are
labelled as follows:
$\vec{x} =
(x_{\nb-1}, x_{\nb-2}, \ldots, x_1, x_0)$,
then
we will say that the components of
$\vec{x}$ are labelled naturally.
For some applications, it is very convenient
to use natural labelling. For other applications,
it doesn't much matter whether we use
natural labelling or not.
In cases where it doesn't matter, we may
use other common labellings such as
$\vec{x} =
(x_1, x_2, \ldots, x_\nb)$.

Let
$\vec{\nu}=
(\nb-1, \nb-2, \ldots, 1,0)$, and
$2^{\vec{\nu}}=(2^{\nb-1}, 2^{\nb-2}, \ldots, 2^1,2^0)$.

Given any $x\in Z_{0, \ns-1}$, we can write
$x=\sum_{i=0}^{\nb-1}2^i x_i$.
If we define the naturally labelled
binary vector $\vec{x} =
(x_{\nb-1}, \ldots, x_1, x_0)$,
then $x=2^{\vec{\nu}}\cdot \vec{x}$.
We call $\vec{x} =
(x_{\nb-1}, \ldots, x_1, x_0)$
 the binary representation
of $x$ and denote it by $bin(x)$.

Given any naturally labelled
binary vector $\vec{x} =
(x_{\nb-1}, \ldots, x_1, x_0)$,
we can write $x=2^{\vec{\nu}}\cdot \vec{x}$.
We call $x\in Z_{0, \ns-1}$ the decimal representation
of $\vec{x}$ and denote it by $dec(\vec{x})$.

If $\vec{x},\vec{y}\in Bool^\nb$,
we will use $\vec{x}\cdot \vec{y}=
 \sum_{i=0}^{\nb-1} x_i y_i$,
 where the addition is normal, not mod 2.

We define the single-qubit states
$\ket{0}$ and $\ket{1}$ by

\beq
\ket{0} =
\left[
\begin{array}{c}
1 \\ 0
\end{array}
\right]
\;\;,\;\;
\ket{1} =
\left[
\begin{array}{c}
0 \\ 1
\end{array}
\right]
\;.
\eeq
Given any
$\vec{x}=(x_1, x_2, \ldots, x_\nb) \in Bool^{\nb}$,
and given a vector of distinct qubit labels
$\vecbitb = (\bitb_1, \bitb_2,
 \ldots, \bitb_\nb)$,
we define the
$\nb$-qubit state
$\ket{\vec{x}}$ as the following tensor product

\beq
\ket{\vec{x}} =
\ket{\vec{x}}_\vecbitb =
\ket{x_1}_{\bitb_1}
\ket{x_2}_{\bitb_2}
\ldots
\ket{x_\nb}_{\bitb_\nb}=
\ket{x_1}\otimes
\ket{x_2}
\ldots\otimes
\ket{x_\nb}
\;.
\label{eq-ket-vec-x-unnatural}
\eeq
For example,

\beq
\ket{01} =
\left[
\begin{array}{c}
1 \\ 0
\end{array}
\right]
\otimes
\left[
\begin{array}{c}
0 \\ 1
\end{array}
\right]
=
\left[
\begin{array}{c}
0 \\ 1 \\ 0 \\0
\end{array}
\right]
\;.
\eeq
With natural labelling,
we would use $\vec{x} =
(x_{\nb-1}, \ldots, x_1, x_0)$,
$\vecbitb=\vec{\nu}$
and $x=\sum_{i=0}^{\nb-1}2^i x_i$.
Instead of
Eq.(\ref{eq-ket-vec-x-unnatural}),
we would have

\beq
\ket{x}=
\ket{\vec{x}} =
\ket{\vec{x}}_{\vec{\nu}} =
\ket{x_{\nb-1}}_{\nb-1}
\ldots
\ket{x_1}_{1}
\ket{x_0}_{0}
=
\ket{x_{\nb-1}}\otimes
\ldots\otimes
\ket{x_1}\otimes
\ket{x_0}
\;.
\label{eq-ket-vec-x-natural}
\eeq

Of course, any $\nb$ qubit state
can be obtained as a linear combination of
the states $\ket{\vec{x}}$
for all $\vec{x}\in Bool^\nb$.

$I_r$ will represent the $r$ dimensional
unit matrix, for any integer $r\geq 1$.

Suppose
$\vecbitb =(\bitb_1, \bitb_2, \ldots, \bitb_\nb)$
is a vector of bit labels,
$(M_1, M_2, \ldots , M_\nb)$ is
a vector of  $2\times 2$ complex matrices,
and
$(\phi_1, \phi_2, \ldots , \phi_\nb)$ is
a vector of  2-dimensional complex column vectors.
For $i\in Z_{1, \nb}$, we define
$M_i(\beta_i)$ by

\beq
M_i(\beta_i) =
I_2 \otimes
\cdots \otimes
I_2 \otimes
M_i \otimes
I_2 \otimes
\cdots \otimes
I_2
\;,
\eeq
where the matrix $M_i$ on the right
hand side is located
at bit position $i$
(counting from left to right, starting at 1)
 in the tensor product
of $\nb$ $2\times 2$ matrices.
We often define a product operator
$M(\vecbitb)$ by

\beq
M(\vecbitb) = \prod_{i=1}^\nb M_i(\beta_i)=
M_1(\bitb_1)\otimes
M_2(\bitb_2)\otimes
\ldots
M_\nb(\bitb_\nb)
\;,
\eeq
and a product state $\ket{\phi}_\vecbitb$

\beq
\ket{\phi}_\vecbitb =
\prod_{i=1}^\nb \ket{\phi_i}_{\beta_i}=
\ket{\phi_1}\otimes
\ket{\phi_2}\otimes
\ldots
\ket{\phi_\nb}
\;.
\eeq
For example, we might find it useful to define
an operator $M(\vecbitb)$
and a state
$\ket{\phi}_{\vecbitb}$
 by

 \beq
M(\vecbitb)
=
\prod_{i=1}^{\nb}\sigx(\bitb_i)=
\sigx\otimes
\sigx\otimes
\ldots\otimes
\sigx
\;,
\eeq

\beq
\ket{\phi}_{\vecbitb}=
\ket{0}_{\vecbitb} =
\prod_{i=1}^{\nb}\ket{0}_{\bitb_i}=
\left(\begin{array}{c}1\\0\end{array}\right)
\otimes
\left(\begin{array}{c}1\\0\end{array}\right)
\otimes\cdots\otimes
\left(\begin{array}{c}1\\0\end{array}\right)
=
[1,0,0,\ldots,0]^T
\;.
\eeq

With natural labelling, we use
$\vecbitb = \vec{\nu}$.
Let
$(M_{\nb-1}, \ldots , M_1, M_0)$
be a vector of
 $2\times 2$ complex matrices,
and let
$(\phi_{\nb-1}, \ldots , \phi_1, \phi_0)$
be a vector of
 2-dimensional complex column vectors.
 With natural labelling,
for $i\in Z_{0, \nb-1}$, we define
$M_i(i)$ by

\beq
M_i(i) =
I_2 \otimes
\cdots \otimes
I_2 \otimes
M_i \otimes
I_2 \otimes
\cdots \otimes
I_2
\;,
\eeq
where the matrix $M_i$ on the right
hand side is located
at bit position $i$
(counting from right to left, starting at 0)
 in the tensor product
of $\nb$ $2\times 2$ matrices.
We often define a product operator
$M(\vec{\nu})$ by

\beq
M(\vec{\nu}) = \prod_{i=0}^{\nb-1} M_i(i)=
M_{\nb-1}(\nb-1)\otimes
\ldots\otimes
M_1(1)\otimes
M_0(0)
\;,
\eeq
and a product state $\ket{\phi}_{\vec{\nu}}$

\beq
\ket{\phi}_{\vec{\nu}} =
\prod_{i=0}^{\nb-1} \ket{\phi_i}_{i}=
\ket{\phi_{\nb-1}}\otimes
\ldots\otimes
\ket{\phi_1}\otimes
\ket{\phi_0}
\;.
\eeq

Next we explain
our circuit diagram notation.
In our
qubit circuit diagrams,
each horizontal wire represents
a single qubit (except when
stated explicitly that the
wire represents several qubits). Different
wires  represent different qubits.
We label single qubit wires
by Greek letters or
by integers as follows:

\beq
\begin{array}{c}
\Qcircuit @C=1em @R=1em @!R{
&\rstick{\bita}\qw
\\
&\rstick{\bitb}\qw
\\
&\rstick{\bitc}\qw
\\
&\rstick{\bitd}\qw
\\
&\vdots
}
\end{array}
\;\;\;\;\;\;,\;\;\;
\begin{array}{c}
\Qcircuit @C=1em @R=1em @!R{
&\rstick{0}\qw
\\
&\rstick{1}\qw
\\
&\rstick{2}\qw
\\
&\rstick{3}\qw
\\
&\vdots
}
\end{array}
\;\;\;\;\;.
\label{eq-qwires}
\eeq
Thus, the first (topmost) wire
is labelled either
$\alpha$ or $0$, the second wire
is labelled either $\beta$ or $1$,
and so forth.
For some special applications,
we label qubits differently from
Eq.(\ref{eq-qwires}). For example,
we might label the first two wires
$\bita_1, \bita_2$, and the next
two wires $\bitb_1, \bitb_2$,
or we might want to label the first
wire $(\bita_1, \bita_2)$,
and make it represent two qubits.
In cases where bit labelling is different
from Eq.(\ref{eq-qwires}),
this will be stated explicitly.
Bras are represented by
\beq
\ket{\psi_1}_\bita
\ket{\psi_2}_\bitb
=
\begin{array}{c}
\Qcircuit @C=1em @R=.5em @!R{
&\gate{\ket{\psi_1}}
\\
&\gate{\ket{\psi_2}}
}
\end{array}
\;\;,\;\;
\ket{\psi}_{\bita\bitb}
=
\begin{array}{c}
\Qcircuit @C=1em @R=1.5em @!R{
&\multigate{1}{\ket{\psi}}
\\
&\ghost{\ket{\psi}}
}
\end{array}
\;,
\eeq
and kets by

\beq
\bra{\chi_1}_\bita
\bra{\chi_2}_\bitb
=
\begin{array}{c}
\Qcircuit @C=1em @R=.5em @!R{
&\freegate{\bra{\chi_1}}&\qw
\\
&\freegate{\bra{\chi_2}}&\qw
}
\end{array}
\;\;,\;\;
\bra{\chi}_{\bita\bitb}
=
\begin{array}{c}
\Qcircuit @C=1em @R=1.5em @!R{
&\freemultigate{1}{\bra{\chi}}&\qw
\\
&\freeghost{\bra{\chi}}&\qw
}
\end{array}
\;.
\eeq
Operators are represented by
\beq
T_1(\bita) T_2(\bitb)
=
\begin{array}{c}
\Qcircuit @C=1em @R=.5em @!R{
&\gate{T_1}&\qw
\\
&\gate{T_2}&\qw
}
\end{array}
\;\;,\;\;
T(\bita,\bitb)
=
\begin{array}{c}
\Qcircuit @C=1em @R=1.5em @!R{
&\multigate{1}{T}&\qw
\\
&\ghost{T}&\qw
}
\end{array}
\;.
\eeq
Matrix elements are represented
by combining the above rules
for bras, kets, and operators.
For example,

\beq
\bra{\chi}_{\bita\bitb}
T(\bita,\bitb)\ket{\psi}_{\bita\bitb}
=
\begin{array}{c}
\Qcircuit @C=1em @R=1.5em @!R{
\freemultigate{1}{\bra{\chi}}
&\multigate{1}{T}
&\multigate{1}{\ket{\psi}}
\\
\freeghost{\bra{\chi}}
&\ghost{T}
&\ghost{\ket{\psi}}
}
\end{array}
\;.
\eeq
{\it Note that in our circuit diagrams,
time flows from the right to the left
of the diagram.} Careful:
Many workers in Quantum
Computing draw their diagrams
so that time flows from
left to right. We eschew their
convention because
it forces one to reverse
the order of the operators
every time one wishes to convert
between a circuit
diagram
and its algebraic equivalent
in Dirac Notation.

Next, we will introduce
a slight enhancement to the standard Dirac
Notation. Given a
ket $\ket{\psi}$,
if we can find an operator
$\Omega$ such that
$\ket{\psi}$ is
a unique (up to a scalar factor) eigenvector
of $\Omega$ with eigenvalue $\lambda$,
then we will sometimes denote
$\ket{\psi}$ by
$\ket{\Omega=\lambda}$.
Sometimes, in order to
specify $\ket{\psi}$ uniquely,
one needs to
find a complete set of commuting operators
$\{\Omega_i : i\in Z_{1, N}\}$
such that $\Omega_i\ket{\psi} =
\lambda_i\ket{\psi}$
for all $i$,
and then we can denote $\ket{\psi}$
by
$\ket{\vec{\Omega}=\vec{\lambda}}$.
Note that if $U$ is a unitary
operator that acts on the same
Hilbert space as an operator $\Omega$, then
$\ket{U\Omega U^\dagger=\lambda}=
U\ket{\Omega=\lambda}$.
If
operator
$\Omega$
has an
eigenspace with eigenvalue $\lambda$,
then we denote the projector onto that
eigenspace by
$\pi(\Omega=\lambda)$.
If the eigenspace is one dimensional, then
$\pi(\Omega=\lambda)=\ket{\Omega=\lambda}
\bra{\Omega=\lambda}$.
If the eigenspace has dimension greater than one,
then we can always find an
orthonormal  basis
$\{\ket{\psi_\lambda^i}: i\in S\}$
for the eigenspace,
and then
$\pi(\Omega=\lambda)=
\sum_{i\in S}
\ket{\psi_\lambda^i}
\bra{\psi_\lambda^i}$.
Note that if $U$ is a unitary
operator
that acts on the same
Hilbert space as operator $\Omega$,
then
$U\pi(\Omega=\lambda)U^\dagger=
\pi(U\Omega U^\dagger=\lambda)$.

The Pauli matrices are defined by:

\beq
\sigx =
\left(
\begin{array}{cc}
0&1\\
1&0
\end{array}
\right)
\;,\;\;
\sigy =
\left(
\begin{array}{cc}
0&-i\\
i&0
\end{array}
\right)\;,\;\;
\sigz =
\left(
\begin{array}{cc}
1&0\\
0&-1
\end{array}
\right)
\;.
\eeq
More information about the Pauli matrices
may be found in the section
entitled Pauli Matrices.

We will often abbreviate
$n-$fold tensor
products of Pauli matrices
as follows. If $w_1, w_2,
\ldots, w_n\in \{X,Y, Z\}$, and
$b_1, b_2,
\ldots, b_n\in Bool$, then let

\beq
\sigma_{w_1,w_2, \ldots, w_n}^{b_1, b_2, \ldots, b_n} =
\sigma_{w_1}^{b_1}\otimes \sigma_{w_2}^{b_2}
\otimes\ldots
\otimes \sigma_{w_n}^{b_n}
\;.
\eeq
For example,
$\sigma_{XYY}^{1,0,1}=
\sigx^1\otimes \sigy^0 \otimes \sigy^1$.
Equivalently, for $n$ bits
$\vecbita=(\alpha_1, \alpha_2,\ldots \alpha_n)$,

\beq
\sigma_{w_1,w_2, \ldots, w_n}^{b_1, b_2, \ldots, b_n}
(\vec{\alpha})=
\prod_{i=1}^n
\sigma_{w_i}^{b_i}(\bita_i)
\;.
\eeq
Also let

\beq
\sigma_{w_1,w_2, \ldots, w_n}=
\sigma_{w_1,w_2, \ldots, w_n}^{1,1, \ldots, 1}=
\sigma_{w_1}\otimes \sigma_{w_2}
\otimes\ldots
\otimes \sigma_{w_n}
\;.
\eeq
For example,
$\sigma_{XYY}=\sigma_{XYY}^{1,1,1}=
\sigx\otimes \sigy \otimes \sigy$.

It is sometimes convenient to define
the following operator for any $x,z\in Bool$
and any qubit $\alpha$:

\beq
\Lambda^{x, z}(\bita)=
\sigx^{x}(\bita)
\sigz^{z}(\bita)
\;.
\eeq
Note that
$\Lambda^{x,z\dagger}=(-1)^{xz}\Lambda^{x, z}$,
and
$\Lambda^{00}=1$,
$\Lambda^{10}=\sigx$,
$\Lambda^{11}=(-i)\sigy$,
$\Lambda^{00}=\sigz$.
$\Lambda^{x, z}$ arises, for example,
when dealing with Bell states.

For any $j\in Bool$ and
$w_1, w_2\in \{X,Y, Z\}$, let
$\Pi_{w_1, w_2}^{j}$ be the projection operator
that projects the
2 qubit Hilbert space onto
the eigenspace of $\sigma_{w_1,w_2}$
with eigenvalue
$(-1)^j$. Thus,

\beq
\Pi_{w_1, w_2}^{j} =
\pi[\sigma_{w_1,w_2}=(-1)^j]
\;.
\eeq
Note that
\beq
\sigma_{ZZ}=
\sigz\otimes \sigz=
\left(
\begin{array}{cc}
1&0\\
0&-1
\end{array}
\right)
\otimes
\left(
\begin{array}{cc}
1&0\\
0&-1
\end{array}
\right)
=
diag(1,-1,-1,1)
\;.
\label{eq-sigzz}
\eeq
From Eq.(\ref{eq-sigzz}),
it is clear that
for any $j,a,b \in Bool$,

\beq
\pizz{j}\ket{a,b}
=\delta^j_{a\oplus b}\ket{a,b}
\;.
\eeq
\section{Pauli Matrices}

 The Pauli matrices are defined by:

\beq
\sigx =
\left(
\begin{array}{cc}
0&1\\
1&0
\end{array}
\right)
\;,\;\;
\sigy =
\left(
\begin{array}{cc}
0&-i\\
i&0
\end{array}
\right)\;,\;\;
\sigz =
\left(
\begin{array}{cc}
1&0\\
0&-1
\end{array}
\right)
\;.
\eeq
Sometimes one refers to
$\sigx,\sigy, \sigz$ as $\sigma_1, \sigma_2, \sigma_3$,
respectively. One can then use $\sigma_0$ to
denote the $2 \times 2$ identity matrix.
It is often convenient to use
the vector of Pauli matrices
$\vec{\sigma} = (\sigx, \sigy, \sigz)$.

All 3 Pauli matrices are their own inverses:
\beq
\sigx^2=\sigy^2=\sigz^2=1
\;.
\label{eq-sig-sq}
\eeq
Distinct Pauli matrices anticommute. For example,

\beq
\sigx\sigx = -\sigy\sigx
\;.
\label{eq-sig-anticom}
\eeq
It is easy to check that

\beq
\sigx\sigy = i \sigz
\;,\;\;
\sigy\sigz = i \sigx
\;,\;\;
\sigz\sigx = i \sigy
\;.
\label{eq-sigx-sigy}
\eeq
Note that Eqs.(\ref{eq-sig-sq}),
(\ref{eq-sig-anticom}) and (\ref{eq-sigx-sigy})
specify a $3 \times 3$ multiplication
table for the 3 Pauli matrices with each other.

For $w\in \{X, Y, Z\}$,
if $\ket{+_w}$ and $\ket{-_w}$ represent
the eigenvectors of
$\sigma_w$ with
eigenvalues $+1$ and $-1$, respectively,
then

\beq
\ket{+_X} =
\frac{1}{\sqrt{2}}
\left(
\begin{array}{c}
1\\1
\end{array}
\right)
\;,\;\;
\ket{-_X} =
\frac{1}{\sqrt{2}}
\left(
\begin{array}{c}
1\\-1
\end{array}
\right)
\;,
\eeq

\beq
\ket{+_Y} =
\frac{1}{\sqrt{2}}
\left(
\begin{array}{c}
1\\i
\end{array}
\right)
\;,\;\;
\ket{-_Y} =
\frac{1}{\sqrt{2}}
\left(
\begin{array}{c}
1\\-i
\end{array}
\right)
\;,
\eeq

\beq
\ket{+_Z} =
\left(
\begin{array}{c}
1\\0
\end{array}
\right)
\;,\;\;
\ket{-_Z} =
\left(
\begin{array}{c}
0\\1
\end{array}
\right)
\;.
\eeq

We define

\beq
\ket{0} =\ket{+_Z}
\;,
\eeq
and

\beq
\ket{1} =\ket{-_Z}
\;.
\eeq
We will use
$n$ to denote the ``number operator". Thus,

\beq
n =  \left(\begin{array}{cc}
0&0\\
0&1
\end{array}\right)
=\ket{-_Z}\bra{-_Z}=  \frac{1-\sigma_Z}{2}
\;,
\eeq
and

\beq
\nbar = 1-n = \left(\begin{array}{cc}
1&0\\
0&0
\end{array}\right)
=\ket{+_Z}\bra{+_Z}=  \frac{1+\sigma_Z}{2}
\;.
\eeq
Since $n$ and $\sigma_Z$ are diagonal, it is
easy to see that

\beq
\sigma_Z =(-1)^n = 1 - 2n
\;.
\eeq

Most of the definitions and results stated so far
for $\sigz$
have counterparts for
$\sigz$ and $\sigy$.
The counterpart results can be
easily proven by applying a rotation that
interchanges the coordinate axes.
Let $w\in \{X, Y, Z\}$.
If $\ket{+_w}$ and $\ket{-_w}$
represent the eigenvectors of $\sigma_w$ with
eigenvalues $+1$ and $-1$, respectively,
then we define

\beq
\ket{0_w} =\ket{+_w}
\;,
\eeq
and

\beq
\ket{1_w} =\ket{-_w}
\;.
\eeq
Let

\beq
n_w=
\ket{-_w}\bra{-_w}=  \frac{1-\sigma_w}{2}
\;,
\eeq

\beq
\nbar_w = 1-n_w
=\ket{+_w}\bra{+_w}=  \frac{1+\sigma_w}{2}
\;.
\eeq
As when $w=Z$, one has

\beq
 \sigma_w = (-1)^{n_w} = 1 - 2n_w
\;.
\eeq

Note that whenever we use
$\ket{0}$, $\ket{1}$ or $n$ ,
without an $X, Y$ or $Z$ subscript,
the subscript $Z$ should be inferred.

The one bit Hadamard matrix is defined by:

\beq
H = \frac{1}{\sqrt{2}}
\left(
\begin{array}{cc}
1&1\\
1&-1
\end{array}
\right)=
\frac{1}{\sqrt{2}}
(\sigx + \sigz)
\;.
\eeq
It is easy to check that

\beq
H^2 =  1
\;,
\eeq

\beq
H\sigx H = \sigz\;,\;\;
H\sigz H = \sigx
\;,
\eeq

\beq
\ket{0_X} =
\frac{\ket{0} + \ket{1}}{\sqrt{2}} =
H \ket{0}
\;,
\eeq

\beq
\ket{1_X} =
\frac{\ket{0} - \ket{1}}{\sqrt{2}} =
H \ket{1}
\;.
\eeq

The matrix $i^n$ is defined by

\beq
i^n =
\left(
\begin{array}{cc}
1&0\\
0&i
\end{array}
\right)
\;.
\eeq
It is easy to check that

\beq
(i^n)^2 = \sigz
\;,
\eeq

\beq
i^n\sigx i^{-n} = \sigy \;,\;\;
i^{-n}\sigx i^{n} = -\sigy
\;.
\eeq

Note that for $a,b\in Bool$,

\beq
\sigx^b\ket{a} = \ket{a\oplus b}
\;,
\eeq

\beq
\sigz^b\ket{a} = (-1)^{ab}\ket{a}
\;,
\eeq

\beq
\bra{a}H\ket{b} = \frac{(-1)^{ab}}{\sqrt{2}}
\;.
\eeq

A general qubit rotation is defined by
$e^{i\vec{\theta}\cdot \vec{\sigma}}$,
where $\vec{\theta}$ is a 3 dimensional
real vector.
For any real
number $\theta$,

\beq
e^{i\theta\sigz} =
\cos\theta + i \sigz \sin\theta
\;.
\label{eq-z-rot}
\eeq
Eq.(\ref{eq-z-rot}) can
be proven by expressing both sides
of it as a
power series.
Applying a rotation to
Eq.(\ref{eq-z-rot}), it becomes

\beq
e^{i\vec{\theta}\cdot \vec{\sigma}}=
\cos\theta +
i\vec{\sigma}\cdot\hat{\theta}
\sin\theta
\;,
\eeq
where $\vec{\theta}$ is a
3 dimensional real vector,
$\theta$ is its magnitude, and
$\hat{\theta}=\vec{\theta}/\theta$.
\section{Hadamard Matrices}

The 1 bit Hadamard matrix is defined by

\beq
H_1 =\frac{1}{\sqrt{2}}
\begin{tabular}{r|rr}
         & {\tiny 0} & {\tiny 1} \\
\hline
{\tiny 0}& 1&  1\\
{\tiny 1}& 1& -1\\
\end{tabular}
\;.
\eeq
The $\nb$-bit Hadamard matrix is defined
as the $\nb$-fold tensor product of $H_1$:

\beq
H_\nb = \underbrace{
H_1\otimes H_1 \otimes \ldots \otimes H_1}_{\mbox{$\nb$ factors}}
\;.
\eeq
For example, for $\nb=2$,

\beq
H_2 =
\frac{1}{2}
\begin{tabular}{r|rrrr}
         & {\tiny 00} & {\tiny 01} & {\tiny 10} & {\tiny 11}\\
\hline
{\tiny 00}&  1&  1&  1&  1\\
{\tiny 01}&  1& -1&  1& -1\\
{\tiny 10}&  1&  1& -1& -1\\
{\tiny 11}&  1& -1& -1&  1\\
\end{tabular}
\;,
\eeq
where we have labelled
the rows and columns with binary numbers
 in increasing dictionary order.
Equivalently, for bits
$\vec{\alpha}=(\alpha_1,
\alpha_1, \ldots, \alpha_{\nb})$,

\beq
H_\nb(\vec{\alpha})=
\prod_{i=1}^{\nb}
H_1(\alpha_i)
\;.
\eeq
We will often use a plain $H$
to represent $H_1$.
Since
$(H_1)_{b,b'} = \frac{(-1)^{bb'}}{\sqrt{2}}$
 for $b, b'\in Bool$,
it follows that

 \beq
 (H_\nb)_{\vec{b},\vec{b'}} =
 \frac{(-1)^{\vec{b}\cdot \vec{b'}}}{\sqrt{2^\nb}}
 \;
 \eeq
 for $\vec{b}, \vec{b'}\in Bool^\nb$ .
Since $H_1^2=1$ and $H_1^T=H_1$, where
T=transpose, it follows that

\beq
H_\nb^2 = 1
\;,
\eeq
and

\beq
H_\nb^T = H_\nb
\;.
\eeq
\section{CNOTs}

We define a CNOT
(C = controlled, NOT $=\sigx$) by:

\grayeq{
\beq
CNOT(\bita\rarrow\bitb)=
CNOT(\bitb\larrow\bita)=
\cnot{\bita}{\bitb}=
(-1)^{n(\bita)n_X(\bitb)}=
\begin{array}{c}
\Qcircuit @C=1em @R=1em @!R{
&\dotgate\qwx[1]
&\qw
\\
&\timesgate
&\qw
}
\end{array}
\;.
\eeq}
$\bita$ is called the
{\bf control qubit} and $\bitb$ is called the
{\bf target qubit}. The CNOT can be easily
generalized to have more than one control qubit:

\grayeq{
\beq
\sigma_X^{n(\bita)n(\bitb)}(\bitc)=
(-1)^{n(\bita)n(\bitb)n_X(\bitc)}=
\begin{array}{c}
\Qcircuit @C=1em @R=1em @!R{
&\dotgate\qwx[2]
&\qw
\\
&\dotgate
&\qw
\\
&\timesgate
&\qw
}
\end{array}
\;.
\eeq}
Other operators related to CNOT are

\grayeq{
\beq
\cnoto{\bita}{\bitb}=
(-1)^{\nbar(\bita)n_X(\bitb)}=
\begin{array}{c}
\Qcircuit @C=1em @R=1em @!R{
&\ogate\qwx[1]
&\qw
\\
&\timesgate
&\qw
}
\end{array}
\;,
\eeq}
and

\grayeq{
\beq
\sigz^{n(\bita)}(\bitb)=
\sigz^{n(\bitb)}(\bita)=
(-1)^{n(\bita)n(\bitb)}=
\begin{array}{c}
\Qcircuit @C=1em @R=1em @!R{
&\dotgate\qwx[1]
&\qw
\\
&\dotgate
&\qw
}
\end{array}
\;.
\eeq}
For any $a,b,c\in Bool$,

\grayeq{
\beq
\cnot{\bita}{\bitb}\ket{a,b}_{\bita\bitb}=
\ket{a, b\oplus a}
\;,
\label{eq-cnot-oplus}
\eeq}

\grayeq{
\beq
\sigma_X^{n(\bita)n(\bitb)}(\bitc)
\ket{a,b,c}_{\bita\bitb\bitc}=
\ket{a, b, c\oplus ab}
\;,
\eeq}

\grayeq{
\beq
\cnoto{\bita}{\bitb}\ket{a,b}_{\bita\bitb}=
\ket{a, b \oplus\overline{a} }
\;,
\eeq}

\grayeq{
\beq
(-1)^{n(\bita)n(\bitb)}\ket{a,b}_{\bita\bitb}=
(-1)^{ab}\ket{a, b}
\;.
\eeq}

Some workers represent a CNOT by
$\begin{array}{c}
\Qcircuit @C=1em @R=.5em @!R{
&\dotgate\qwx[1]
&\qw
\\
&\targ
&\qw
}
\end{array}$
instead of
$\begin{array}{c}
\Qcircuit @C=1em @R=1em @!R{
&\dotgate\qwx[1]
&\qw
\\
&\timesgate
&\qw
}
\end{array}$.
The
$\begin{array}{c}
\Qcircuit @C=1em @R=.5em @!R{
&\dotgate\qwx[1]
&\qw
\\
&\targ
&\qw
}
\end{array}$
notation reminds us of the $\oplus$
in Eq.(\ref{eq-cnot-oplus}), whereas the
$\begin{array}{c}
\Qcircuit @C=1em @R=1em @!R{
&\dotgate\qwx[1]
&\qw
\\
&\timesgate
&\qw
}
\end{array}$
notation reminds us
of the $X$ in $\cnot{\bita}{\bitb}$.

\claim

\grayeq{
\beq
\cnot{\bitb}{\bita}=
\sigx(\bita)n(\bitb) + \nbar(\bitb)
\;.
\eeq}
\proof
Check that both sides agree when $n(\bitb)$
equals zero and one.
\qed

\claim

\grayeq{
\beq
\cnot{\bitb}{\bita}=
\frac{1}{2}\sum_{(x,z)\in Bool^2}
\sigx^x(\bita)\sigz^z(\bitb) (-1)^{xz}
\;.
\eeq}
\proof
\beqa
\cnot{\bitb}{\bita} &=&
(-1)^{n_X(\bita)n_Z(\bitb)}\\
&=& 1 - 2 n_X(\bita)n_Z(\bitb)\\
&=& 1 - 2
\left(\frac{1-\sigx(\bita)}{2}\right)
\left(\frac{1-\sigz(\bitb)}{2}\right)\\
&=&
\frac{1}{2}
[ 1 + \sigx(\bita) + \sigz(\bitb)
- \sigma_{XZ}(\bita, \bitb)]
\;.
\eeqa
\qed

\claim (Permuting 2 CNOTs in a chain)

\grayeq{
\beqa
\begin{array}{c}
\Qcircuit @C=1em @R=1em @!R{
&\timesgate\qwx[1]
&\qw
&\qw
\\
&\dotgate
&\timesgate\qwx[1]
&\qw
\\
&\qw
&\dotgate
&\qw
}
\end{array}
&=&
\begin{array}{c}
\Qcircuit @C=1em @R=1em @!R{
&\timesgate\qwx[2]
&\qw
&\timesgate\qwx[1]
&\qw
\\
&\qw
&\timesgate\qwx[1]
&\dotgate
&\qw
\\
&\dotgate
&\dotgate
&\qw
&\qw
\gategroup{1}{2}{3}{2}{.7em}{.}
}
\end{array}
\label{eq-com-cnots-line}
\\
&=&
\begin{array}{c}
\Qcircuit @C=1em @R=1em @!R{
&\qw
&\timesgate\qwx[1]
&\timesgate\qwx[2]
&\qw
\\
&\timesgate\qwx[1]
&\dotgate
&\qw
&\qw
\\
&\dotgate
&\qw
&\dotgate
&\qw
\gategroup{1}{4}{3}{4}{.7em}{.}
}
\end{array}
\;.
\eeqa}
\proof
Let LHS and RHS stand for the left and
right hand sides of Eq.(\ref{eq-com-cnots-line}).
For $a,b,c\in Bool$,

\beqa
LHS\ket{a,b,c}_{\bita\bitb\bitc}&=&
\cnot{\bitb}{\bita}
\cnot{\bitc}{\bitb}
\ket{a,b,c}\\
&=&
\cnot{\bitb}{\bita}\ket{a,b\oplus c,c}\\
&=&
\ket{a\oplus b \oplus c,b\oplus c,c}
\;.
\eeqa

\beqa
RHS\ket{a,b,c}_{\bita\bitb\bitc}&=&
\cnot{\bitc}{\bita}
\cnot{\bitc}{\bitb}
\cnot{\bitb}{\bita}
\ket{a,b,c} \\
&=&
\cnot{\bitc}{\bita}
\cnot{\bitc}{\bitb}
\ket{a\oplus b,b,c}\\
&=&
\cnot{\bitc}{\bita}
\ket{a\oplus b,b \oplus c,c}\\
&=&
\ket{a\oplus b \oplus c,b \oplus c,c}
\;.
\eeqa

Finally, note that
$CNOT(\bitc\rarrow\bita)$ and
$CNOT(\bitc\rarrow\bitb)
CNOT(\bitb\rarrow\bita)$ commute.
\qed

A mnemonic for remembering
Eq.(\ref{eq-com-cnots-line}):
On the left hand side of
Eq.(\ref{eq-com-cnots-line}), we have a
``chain"
CNOT($\bita\larrow\bitb$)
CNOT($\bitb\larrow\bitc$) of CNOTs.
When CNOT($\bita\larrow\bitb$) is moved
to the right (or to the left), over
CNOT($\bitb\larrow\bitc$), it leaves
behind as a ``wake"
the CNOT within the dotted box.
The wake CNOT($\bita\larrow\bitc$)
points from the beginning to the
end of the  original chain
CNOT($\bita\larrow\bitb$)
CNOT($\bitb\larrow\bitc$).

Throughout QC Paulinesia,
we will refer to equations,
like Eq.(\ref{eq-com-cnots-line}),
wherein two operators are permuted
and a wake is produced,
as ``wake identities".
Eq.(\ref{eq-com-cnots-line})
is the first of many
wake identities we will present.

\claim (Permuting 2 CNOTs in a chain,
when first and
last qubit of chain are the same)

\grayeq{
\beq
\begin{array}{c}
\Qcircuit @C=1em @R=1em @!R{
&\timesgate\qwx[1]
&\dotgate
&\qw
\\
&\dotgate
&\timesgate\qwx[-1]
&\qw
}
\end{array}
=
\begin{array}{c}
\Qcircuit @C=1em @R=1em @!R{
&\dotgate
&\timesgate\qwx[1]
&\dotgate
&\timesgate\qwx[1]
&\qw
\\
&\timesgate\qwx[-1]
&\dotgate
&\timesgate\qwx[-1]
&\dotgate
&\qw
\gategroup{2}{3}{1}{2}{.7em}{.}
}
\end{array}
\;.
\label{eq-com-cnots-loop}
\eeq}
\proof
Eq.(\ref{eq-com-cnots-loop}) is the same as

\beq
1=
\begin{array}{c}
\Qcircuit @C=1em @R=1em @!R{
&\timesgate\qwx[1]
&\dotgate
&\timesgate\qwx[1]
&\dotgate
&\timesgate\qwx[1]
&\dotgate
&\qw
\\
&\dotgate
&\timesgate\qwx[-1]
&\dotgate
&\timesgate\qwx[-1]
&\dotgate
&\timesgate\qwx[-1]
&\qw
}
\end{array}
\;,
\eeq
which is just the fact
that $E^2=1$, where $E$ is
the exchange operator.
\qed

A mnemonic for remembering
Eq.(\ref{eq-com-cnots-loop}):
On the left hand side of
Eq.(\ref{eq-com-cnots-loop}), we have a
``loop chain"
CNOT($\bita\larrow\bitb$)
CNOT($\bitb\larrow\bita$) of CNOTs.
When CNOT($\bita\larrow\bitb$) is moved over
CNOT($\bitb\larrow\bita$), it leaves
behind as a ``wake"
the two CNOTs within the dotted box.
The wake and the non-wake parts are
identical.

\claim

\grayeq{
\beq
\begin{array}{c}
\Qcircuit @C=1em @R=1em @!R{
&\dotgate\qwx[1]
&\qw
\\
&\timesgate
&\gate{\sigz}
}
\end{array}
=
\begin{array}{c}
\Qcircuit @C=1em @R=1em @!R{
&\gate{\sigz}
&\qw
&\dotgate\qwx[1]
&\qw
\\
&\qw
&\gate{\sigz}
&\timesgate
&\qw
\gategroup{1}{2}{1}{2}{.7em}{.}
}
\end{array}
\;.
\label{eq-perm-cnot-sigz}
\eeq}
(Dotted box encloses wake.)
\proof
Let LHS and RHS stand for the
left and right hand sides of
Eq.(\ref{eq-perm-cnot-sigz}).
For $a,b\in Bool$,

\beqa
LHS\ket{a,b}_{\bita\bitb}&=&
\cnot{\bita}{\bitb}\sigz(\bitb)\ket{a,b}
\\
&=& (-1)^b \ket{a, b\oplus a}
\;.
\eeqa

\beqa
RHS\ket{a,b}_{\bita\bitb}&=&
\sigz(\bita)\sigz(\bitb)
\cnot{\bita}{\bitb}\ket{a,b}
\\
&=&
\sigz(\bita)\sigz(\bitb)
\ket{a, b\oplus a}
\\
&=& (-1)^b \ket{a, b\oplus a}
\;.
\eeqa
\altproof

\beqa
\cnot{\bita}{\bitb}
\sigz(\bitb)
\cnot{\bita}{\bitb}
&=&
[\sigx(\bitb) n(\bita) + \nbar(\bita)]
\sigz(\bitb)
[\sigx(\bitb) n(\bita) + \nbar(\bita)]\\
&=&
\sigz(\bitb)
[-\sigx(\bitb) n(\bita) + \nbar(\bita)]
[\sigx(\bitb) n(\bita) + \nbar(\bita)]\\
&=&
\sigz(\bitb)[-n(\bita) + \nbar(\bita)]\\
&=&
\sigz(\bitb)\sigz(\bita)
\;.
\eeqa
\qed

\claim

\grayeq{
\beq
\begin{array}{c}
\Qcircuit @C=1em @R=1em @!R{
&\dotgate\qwx[1]
&\qw
&\dotgate\qwx[1]
&\qw
\\
&\timesgate
&\dotgate\qwx[1]
&\timesgate
&\qw
\\
&\qw
&\timesgate
&\qw
&\qw
}
\end{array}
=
\begin{array}{c}
\Qcircuit @C=1em @R=1em @!R{
&\dotgate\qwx[2]
&\qw
&\qw
\\
&\qw
&\dotgate\qwx[1]
&\qw
\\
&\timesgate
&\timesgate
&\qw
}
\end{array}
\;.
\label{eq-two-brothers}
\eeq}
\proof
Apply Eq.(\ref{eq-com-cnots-line}) once to
left hand side of Eq.(\ref{eq-two-brothers}).
\qed

Note that in Eq.(\ref{eq-two-brothers}),
the left hand side contains only nearest
neighbor CNOTs,
whereas
the right hand side
contains only commuting CNOTs.

\claim

\grayeq{
\beq
\begin{array}{c}
\Qcircuit @C=1em @R=1em @!R{
&\dotgate\qwx[1]
&\qw
&\qw
&\qw
&\dotgate\qwx[1]
&\qw
\\
&\timesgate
&\dotgate\qwx[1]
&\qw
&\dotgate\qwx[1]
&\timesgate
&\qw
\\
&\qw
&\timesgate
&\dotgate\qwx[1]
&\timesgate
&\qw
&\qw
\\
&\qw
&\qw
&\timesgate
&\qw
&\qw
&\qw
}
\end{array}
=
\begin{array}{c}
\Qcircuit @C=1em @R=1em @!R{
&\dotgate\qwx[3]
&\qw
&\qw
&\qw
\\
&\qw
&\dotgate\qwx[2]
&\qw
&\qw
\\
&\qw
&\qw
&\dotgate\qwx[1]
&\qw
\\
&\timesgate
&\timesgate
&\timesgate
&\qw
}
\end{array}
\;.
\label{eq-three-brothers}
\eeq}
\proof
Apply Eq.(\ref{eq-com-cnots-line}) twice to
left hand side of Eq.(\ref{eq-three-brothers}).
\qed

\claim

\grayeq{
\beq
\begin{array}{c}
\Qcircuit @C=1em @R=1em @!R{
&\dotgate\qwx[1]
&\qw
&\dotgate\qwx[1]
&\qw
&\qw
\\
&\timesgate
&\dotgate\qwx[1]
&\timesgate
&\dotgate\qwx[1]
&\qw
\\
&\qw
&\timesgate
&\qw
&\timesgate
&\qw
}
\end{array}
=
\begin{array}{c}
\Qcircuit @C=1em @R=1em @!R{
&\dotgate\qwx[2]
&\qw
\\
&\qw
&\qw
\\
&\timesgate
&\qw
}
\end{array}
\;.
\label{eq-cnot-next-nearest}
\eeq}
\proof
This follows immediately
from Eq.(\ref{eq-two-brothers}).
\qed

\claim

\grayeq{
\beq
\begin{array}{c}
\Qcircuit @C=1em @R=1em @!R{
&\dotgate\qwx[1]
&\qw
&\qw
&\qw
&\dotgate\qwx[1]
&\qw
&\qw
&\qw
&\qw
\\
&\timesgate
&\dotgate\qwx[1]
&\qw
&\dotgate\qwx[1]
&\timesgate
&\dotgate\qwx[1]
&\qw
&\dotgate\qwx[1]
&\qw
\\
&\qw
&\timesgate
&\dotgate\qwx[1]
&\timesgate
&\qw
&\timesgate
&\dotgate\qwx[1]
&\timesgate
&\qw
\\
&\qw
&\qw
&\timesgate
&\qw
&\qw
&\qw
&\timesgate
&\qw
&\qw
}
\end{array}
=
\begin{array}{c}
\Qcircuit @C=1em @R=1em @!R{
&\dotgate\qwx[3]
&\qw
\\
&\qw
&\qw
\\
&\qw
&\qw
\\
&\timesgate
&\qw
}
\end{array}
\;.
\label{eq-cnot-next-next-nearest}
\eeq}
\proof
The product of left hand sides
of Eqs.(\ref{eq-two-brothers})
and (\ref{eq-three-brothers}),
equals the product of their
right hand sides.
\qed

Eqs.(\ref{eq-cnot-next-nearest}) and
(\ref{eq-cnot-next-next-nearest})
suggest a way of converting a
non-nearest neighbor CNOT
into a sequence
of nearest neighbor ones.

\section{CNOT Generalizations}

In this section, $\vec{\bita}$,
$\vec{\bitb}$ and $\vec{\bitc}$
will denote disjoint
sets of distinct qubits.
That is,
any two different components of the same
vector, or two components of
different vectors
represent different qubits.

Suppose $U$ is a unitary matrix.
Furthermore,
for $j=1,2$, suppose
$\pi_j$
is a projection operator
(i.e., $\pi_j^2 = \pi_j$,
the eigenvalues of $\pi_j$
are all 0 or 1).
Some examples of projection operators
$\pi_j$ that
are of interest to us: 1, $n(\bita)$,
$n(\bita)n(\bitb)$,
$n(\bita)\nbar(\bitb)$,
$n(\bita)n(\bitb)n(\bitc)$, etc.
It is convenient to generalize CNOT
diagrammatic notation as follows.
Let

\grayeq{
\beq
\begin{array}{c}
\Qcircuit @C=1em @R=1em @!R{
&\ovalgate{\pi_1}\qwx[1]
&\rstick{\vec{\bita}}\qw
\\
&\ovalgate{\pi_2}
&\rstick{\vec{\bitb}}\qw
}
\end{array}
\;\;\;=
(-1)^{\pi_1(\vec{\bita})\pi_2(\vec{\bitb})}
\;,
\label{eq-oval-oval}
\eeq}
and

\grayeq{
\beq
\begin{array}{c}
\Qcircuit @C=1em @R=1em @!R{
&\ovalgate{\pi_1}\qwx[1]
&\rstick{\vec{\bita}}\qw
\\
&\gate{U}
&\rstick{\vec{\bitb}}\qw
}
\end{array}
\;\;\;=
{U(\vec{\bitb})}^{\pi_1(\vec{\bita})}
\;.
\label{eq-oval-sq}
\eeq}
We will refer to an operator of the form
Eq.(\ref{eq-oval-sq})
as a {\bf projector controlled unitary operator},
or simply as a {\bf controlled U},
in analogy to a controlled NOT,
for which $U=\sigx=$ the NOT operator.
The set of operators of the form
Eq.(\ref{eq-oval-oval})
is a subset of the set of operators of the form
Eq.(\ref{eq-oval-sq}). Indeed,
given any projection operator
$\pi_2(\vec{\bitb})$, one can always define
the unitary operator
$U(\vec{\bitb})
 = (-1)^{\pi_2(\vec{\bitb})}
= 1 - 2\pi_2(\vec{\bitb})$. Hence,

\beq
\begin{array}{c}
\Qcircuit @C=1em @R=1em @!R{
&\ovalgate{\pi_1}\qwx[1]
&\qw
\\
&\gate{(-1)^{\pi_2}}
&\qw
}
\end{array}
=
\begin{array}{c}
\Qcircuit @C=1em @R=1em @!R{
&\ovalgate{\pi_1}\qwx[1]
&\qw
\\
&\ovalgate{\pi_2}
&\qw
}
\end{array}
\;.
\eeq
Special cases of
Eqs.(\ref{eq-oval-oval}) and
(\ref{eq-oval-sq}) are:

\beq
(-1)^{n(\bita)n(\bitb)}
=
\begin{array}{c}
\Qcircuit @C=1em @R=1em @!R{
&\dotgate\qwx[1]
&\qw
\\
&\dotgate
&\qw
}
\end{array}
=
\begin{array}{c}
\Qcircuit @C=1em @R=1em @!R{
&\ovalgate{n}\qwx[1]
&\qw
\\
&\ovalgate{n}
&\qw
}
\end{array}
=
\begin{array}{c}
\Qcircuit @C=1em @R=1em @!R{
&\ovalgate{n}\qwx[1]
&\qw
\\
&\gate{\sigz}
&\qw
}
\end{array}
\;,
\eeq

\beq
\cnot{\bita}{\bitb}
=
\begin{array}{c}
\Qcircuit @C=1em @R=1em @!R{
&\dotgate\qwx[1]
&\qw
\\
&\timesgate
&\qw
}
\end{array}
=
\begin{array}{c}
\Qcircuit @C=1em @R=1em @!R{
&\ovalgate{n}\qwx[1]
&\qw
\\
&\ovalgate{n_X}
&\qw
}
\end{array}
=
\begin{array}{c}
\Qcircuit @C=1em @R=1em @!R{
&\ovalgate{n}\qwx[1]
&\qw
\\
&\gate{\sigx}
&\qw
}
\end{array}
\;,
\eeq
and, for any $2\times 2$ unitary matrix $U$:

\beq
U(\bitb)^{n(\bita)}
=
\begin{array}{c}
\Qcircuit @C=1em @R=1em @!R{
&\dotgate\qwx[1]
&\qw
\\
&\gate{U}
&\qw
}
\end{array}
=
\begin{array}{c}
\Qcircuit @C=1em @R=1em @!R{
&\ovalgate{n}\qwx[1]
&\qw
\\
&\gate{U}
&\qw
}
\end{array}
\;,
\label{eq-n-one-cu}
\eeq
\beq
U(\bitc)^{n(\bita)n(\bitb)}
=
\begin{array}{c}
\Qcircuit @C=1em @R=.5em @!R{
&\dotgate\qwx[2]
&\qw
\\
&\dotgate
&\qw
\\
&\gate{U}
&\qw
}
\end{array}
=
\begin{array}{c}
\Qcircuit @C=1em @R=.5em @!R{
&\ovalgate{n}\qwx[1]
&\qw
\\
&\ovalgate{n}\qwx[1]
&\qw
\\
&\gate{U}
&\qw
}
\end{array}
\;.
\label{eq-n-two-cu}
\eeq
We will refer to the
operator of Eq.(\ref{eq-n-one-cu})
as an $n^1$ {\bf controlled U},
and to the operator of
Eq.(\ref{eq-n-two-cu})
as an $n^2$ {\bf controlled U}.

Suppose $U$ is any $2\times 2$ unitary matrix.
It can always be diagonalized as follows:

\beq
U = V diag(
e^{i\theta_1},
e^{i\theta_2})
V^\dagger
\;,
\eeq
where $\theta_1, \theta_2$
are reals numbers and
$V$ is a unitary matrix. If we set

\beq
\Delta = \frac{\theta_1-\theta_2}{2}
\;,
\eeq
and

\beq
\overline{\theta} = \frac{\theta_1+\theta_2}{2}
\;,
\eeq
then

\beq
U= e^{i\overline{\theta}}
V e^{i\Delta \sigz}
V^\dagger
\;.
\label{eq-u-eigen-decomp}
\eeq

\claim

For any $2\times 2$ unitary matrix
$U(\bitb)$ given by Eq.(\ref{eq-u-eigen-decomp}),
and projection operator $\pi_1(\vecbita)$,

\grayeq{
\beq
\begin{array}{c}
\Qcircuit @C=1em @R=1em @!R{
&\ovalgate{\pi_1}\qwx[1]
&\qw
\\
&\gate{U}
&\qw
}
\end{array}
=
\begin{array}{c}
\Qcircuit @C=1em @R=1em @!R{
&\gate{e^{i\overline{\theta}\pi_1}}
&\qw
&\ovalgate{\pi_1}\qwx[1]
&\qw
&\ovalgate{\pi_1}\qwx[1]
&\qw
&\rstick{\vec{\bita}}\qw
\\
&\gate{V}
&\gate{e^{i\frac{\Delta}{2}\sigz}}
&\timesgate
&\gate{e^{-i\frac{\Delta}{2}\sigz}}
&\timesgate
&\gate{V^\dagger}
&\rstick{\bitb}\qw
}
\end{array}
\;\;\;\;.
\label{eq-pi-contr-u-decompo}
\eeq}
\proof
Check that both sides agree when $\pi_1$
equals 0 and 1.
\altproof
\beq
U(\bitb)^{n(\bita)}=
e^{i\overline{\theta}n(\bita)}
V(\bitb) e^{i\Delta \sigz(\bitb)n(\bita)}
V(\bitb)^\dagger
\;.
\eeq

\beqa
e^{i\Delta \sigz(\bitb)n(\bita)}
&=&
e^{i\Delta \sigz(\bitb)
\frac{1}{2}[1-\sigz(\bita)]}
\\
&=&
e^{i \frac{\Delta}{2}\sigz(\bitb)}
e^{-i \frac{\Delta}{2}
\sigz(\bitb)
\sigz(\bita)}
\\
&=&
e^{i \frac{\Delta}{2}\sigz(\bitb)}
\sigx(\bitb)^{n(\bita)}
e^{-i \frac{\Delta}{2}
\sigz(\bitb)}
\sigx(\bitb)^{n(\bita)}
\;.
\eeqa
This proof still holds if we  replace $n(\bita)$
by $\pi_1(\vec{\bita})$ and $\sigz(\bita)$
by $(-1)^{\pi_1(\vec{\bita})}$.
\qed

Examples of Eq.(\ref{eq-pi-contr-u-decompo})
are:

\grayeq{
\beq
\begin{array}{c}
\Qcircuit @C=1em @R=.5em @!R{
&\dotgate\qwx[1]
&\qw
\\
&\gate{U}
&\qw
}
\end{array}
=
\begin{array}{c}
\Qcircuit @C=1em @R=.5em @!R{
&\gate{e^{i\bar{\theta}n}}
&\qw
&\dotgate\qwx[1]
&\qw
&\dotgate\qwx[1]
&\qw
&\qw
\\
&\gate{V}
&\gate{e^{i\frac{\Delta}{2}\sigz}}
&\timesgate
&\gate{e^{-i\frac{\Delta}{2}\sigz}}
&\timesgate
&\gate{V^\dagger}
&\qw
}
\end{array}
\;,
\label{eq-n-one-contr-u}
\eeq}
and

\grayeq{
\beq
\begin{array}{c}
\Qcircuit @C=1em @R=.5em @!R{
&\dotgate\qwx[1]
&\qw
\\
&\dotgate\qwx[1]
&\qw
\\
&\gate{U}
&\qw
}
\end{array}
=
\begin{array}{c}
\Qcircuit @C=1em @R=.5em @!R{
&\dotgate\qwx[1]
&\qw
&\dotgate\qwx[1]
&\qw
&\dotgate\qwx[1]
&\qw
&\qw
\\
&\gate{e^{i\bar{\theta}n}}
&\qw
&\dotgate\qwx[1]
&\qw
&\dotgate\qwx[1]
&\qw
&\qw
\\
&\gate{V}
&\gate{e^{i\frac{\Delta}{2}\sigz}}
&\timesgate
&\gate{e^{-i\frac{\Delta}{2}\sigz}}
&\timesgate
&\gate{V^\dagger}
&\qw
}
\end{array}
\;.
\label{eq-n-two-contr-u}
\eeq}
Eqs.(\ref{eq-n-one-contr-u}) and
(\ref{eq-n-two-contr-u}) suggest a way
of converting any $n^r$ controlled
$U$, for an integer $r\geq 1$,
into a sequence of gates containing no
controlled $U$'s but containing
$n^s$ controlled NOTs,
where $s\leq r$.

\claim(Permuting two projector
controlled $U$'s)

Suppose $\pi_1(\vecbita), \pi_2(\vecbita)$
are  commuting
($[\pi_1, \pi_2]=0$)
projection operators
and $U_1(\vecbitb), U_2(\vecbitb)$
are unitary operators. Then

\grayeq{
\beq
\begin{array}{c}
\Qcircuit @C=1em @R=.5em @!R{
&\ovalgate{\pi_1}\qwx[1]
&\ovalgate{\pi_2}\qwx[1]
&\qw
\\
&\gate{U_1}
&\gate{U_2}
&\qw
}
\end{array}
=
\begin{array}{c}
\Qcircuit @C=1em @R=.5em @!R{
&\ovalgate{\pi_1\pi_2}\qwx[1]
&\ovalgate{\pi_2}\qwx[1]
&\ovalgate{\pi_1}\qwx[1]
&\rstick{\vecbita}\qw
\\
&\gate{U_1 U_2 U_1^\dagger U_2^\dagger}
&\gate{U_2}
&\gate{U_1}
&\rstick{\vecbitb}\qw
\gategroup{1}{2}{2}{2}{.7em}{.}
}
\end{array}
\;\;\;\;.
\label{eq-perm-two-pi-contr-u}
\eeq}
(Dotted box encloses wake.)
Algebraically,

\beq
U_1(\vecbitb)^{\pi_1(\vecbita)}
U_2(\vecbitb)^{\pi_2(\vecbita)}
=
(U_1 U_2 U^\dagger_1 U^\dagger_2)^{\pi_1\pi_2}
U_2^{\pi_2}U_1^{\pi_1}
\;.
\eeq
\proof
Check that both sides of
Eq.(\ref{eq-perm-two-pi-contr-u})
agree when $(\pi_1,\pi_2)$
equals each element of $Bool^2$.
\qed

\claim

For any projection operator
$\pi_1(\vec{\bita})$ and
unitary matrix $U(\vec{\bitc})$,

\grayeq{
\beq
\begin{array}{c}
\Qcircuit @C=1em @R=.5em @!R{
&\ovalgate{\pi_1}\qwx[1]
&\qw
&\qw
\\
&\timesgate
&\dotgate\qwx[1]
&\qw
\\
&\qw
&\gate{U}
&\qw
}
\end{array}
=
\begin{array}{c}
\Qcircuit @C=1em @R=.5em @!R{
&\ovalgate{\pi_1}\qwx[2]
&\ovalgate{\pi_1}\qwx[2]
&\qw
&\ovalgate{\pi_1}\qwx[1]
&\rstick{\vec{\bita}}\qw
\\
&\dotgate
&\qw
&\dotgate\qwx[1]
&\timesgate
&\rstick{\bitb}\qw
\\
&\gate{U^{-2}}
&\gate{U}
&\gate{U}
&\qw
&\rstick{\vec{\bitc}}\qw
\gategroup{1}{2}{3}{3}{.7em}{.}
}
\end{array}
\;\;\;\;.
\label{eq-perm-gen-times-dot}
\eeq}
(Dotted box encloses wake.)
\proof
Consider Eq.(\ref{eq-perm-two-pi-contr-u})
with the following replacements:
$U_1\rarrow \sigx(\bitb)$,
$U_2\rarrow U(\vecbitc)^{n(\bitb)}$,
$\pi_2\rarrow 1$.
Thus,

\beq
U_1 U_2 U_1^\dagger U_2^\dagger
\rarrow
\sigx(\bitb)
U(\vecbitc)^{n(\bitb)}
\sigx(\bitb)
U(\vecbitc)^{-n(\bitb)}
=U(\vecbitc)^{\nbar(\bitb)-n(\bitb)}
=U(\vecbitc)^{1-2n(\bitb)}
\;.
\eeq
\qed

\claim

For any projection operator
$\pi_1(\vec{\bita})$ and
unitary matrix $U(\vec{\bitc})$,

\grayeq{
\beq
\begin{array}{c}
\Qcircuit @C=1em @R=.5em @!R{
&\ovalgate{\pi_1}\qwx[2]
&\qw
\\
&\dotgate
&\qw
\\
&\gate{U}
&\qw
}
\end{array}
=
\begin{array}{c}
\Qcircuit @C=1em @R=.5em @!R{
&\ovalgate{\pi_1}\qwx[2]
&\ovalgate{\pi_1}\qwx[1]
&\qw
&\ovalgate{\pi_1}\qwx[1]
&\qw
&\rstick{\vec{\bita}}\qw
\\
&\qw
&\timesgate
&\dotgate\qwx[1]
&\timesgate
&\dotgate\qwx[1]
&\rstick{\bitb}\qw
\\
&\gate{U^{\frac{1}{2}}}
&\qw
&\gate{U^{\frac{-1}{2}}}
&\qw
&\gate{U^{\frac{1}{2}}}
&\rstick{\vec{\bitc}}\qw
}
\end{array}
\;\;\;.
\label{eq-pi-n-contr-u}
\eeq}
\proof
Apply Eq.(\ref{eq-perm-gen-times-dot})
to the right hand side of
Eq.(\ref{eq-pi-n-contr-u}) to permute
$\sigx(\bitb)^{\pi_1(\vecbita)}$
and
$U(\vecbitc)^{\frac{-1}{2}n(\bitb)}$.
\qed

Examples of Eq.(\ref{eq-pi-n-contr-u}) are

\grayeq{
\beq
\begin{array}{c}
\Qcircuit @C=1em @R=.5em @!R{
&\dotgate\qwx[2]
&\qw
\\
&\dotgate
&\qw
\\
&\gate{U}
&\qw
}
\end{array}
=
\begin{array}{c}
\Qcircuit @C=1em @R=.5em @!R{
&\dotgate\qwx[2]
&\dotgate\qwx[1]
&\qw
&\dotgate\qwx[1]
&\qw
&\qw
\\
&\qw
&\timesgate
&\dotgate\qwx[1]
&\timesgate
&\dotgate\qwx[1]
&\qw
\\
&\gate{U^{\frac{1}{2}}}
&\qw
&\gate{U^{\frac{-1}{2}}}
&\qw
&\gate{U^{\frac{1}{2}}}
&\qw
}
\end{array}
\;,
\label{eq-n-two-sq-root-u}
\eeq}
and

\grayeq{
\beq
\begin{array}{c}
\Qcircuit @C=1em @R=.5em @!R{
&\dotgate\qwx[1]
&\qw
\\
&\dotgate\qwx[2]
&\qw
\\
&\dotgate
&\qw
\\
&\gate{U}
&\qw
}
\end{array}
=
\begin{array}{c}
\Qcircuit @C=1em @R=.5em @!R{
&\dotgate\qwx[1]
&\dotgate\qwx[1]
&\qw
&\dotgate\qwx[1]
&\qw
&\qw
\\
&\dotgate\qwx[2]
&\dotgate\qwx[1]
&\qw
&\dotgate\qwx[1]
&\qw
&\qw
\\
&\qw
&\timesgate
&\dotgate\qwx[1]
&\timesgate
&\dotgate\qwx[1]
&\qw
\\
&\gate{U^{\frac{1}{2}}}
&\qw
&\gate{U^{\frac{-1}{2}}}
&\qw
&\gate{U^{\frac{1}{2}}}
&\qw
}
\end{array}
\;.
\label{eq-n-three-sq-root-u}
\eeq}

Eqs.(\ref{eq-n-two-sq-root-u})
and (\ref{eq-n-three-sq-root-u})
suggest a way of converting
an $n^r$ controlled $U$,
for an integer $r\geq 2$,
into a sequence of gates
that contains no controlled $U$'s
except $n^1$ controlled $U$'s.

\claim

Suppose $\pi_1(\vecbita)$
and $\pi_2(\vecbita)$ are
commuting projection operators. Then

\grayeq{
\beq
\begin{array}{c}
\Qcircuit @C=1em @R=1em @!R{
&\ovalgate{\pi_1}\qwx[1]
&\ovalgate{\pi_2}\qwx[1]
&\qw
\\
&\timesgate
&\dotgate
&\qw
}
\end{array}
=
\begin{array}{c}
\Qcircuit @C=1em @R=1em @!R{
&\gate{(-1)^{\pi_1\pi_2}}
&\ovalgate{\pi_2}\qwx[1]
&\ovalgate{\pi_1}\qwx[1]
&\rstick{\vec{\bita}}\qw
\\
&\qw
&\dotgate
&\timesgate
&\rstick{\bitb}\qw
\gategroup{1}{2}{1}{2}{.7em}{.}
}
\end{array}
\;\;\;.
\label{eq-perm-gen-chain}
\eeq}
(Dotted box encloses wake.)
\proof
Consider Eq.(\ref{eq-perm-two-pi-contr-u})
with the following replacements:
$U_1\rarrow \sigx(\bitb)$,
$U_2\rarrow \sigz(\bitb)$.
Thus,

\beq
U_1 U_2 U_1^\dagger U_2^\dagger
\rarrow
\sigx\sigz\sigx\sigz = -1
\;.
\eeq
\qed

Eq.(\ref{eq-perm-gen-chain})
can be used to transform
sequences of $n^r$ controlled
NOTs. For example,
the following identity
can be easily proven by applying
Eq.(\ref{eq-perm-gen-chain}):

\beq
\begin{array}{c}
\Qcircuit @C=1em @R=1em @!R{
&\qw
&\dotgate\qwx[4]
&\qw
&\dotgate\qwx[4]
&\qw
\\
&\dotgate\qwx[2]
&\qw
&\dotgate\qwx[2]
&\qw
&\qw
\\
&\dotgate
&\qw
&\dotgate
&\qw
&\qw
\\
&\timesgate
&\dotgate
&\timesgate
&\dotgate
&\qw
\\
&\qw
&\timesgate
&\qw
&\timesgate
&\qw
}
\end{array}
=
\begin{array}{c}
\Qcircuit @C=1em @R=1em @!R{
&\dotgate\qwx[4]
&\qw
\\
&\dotgate
&\qw
\\
&\dotgate
&\qw
\\
&\qw
&\qw
\\
&\timesgate
&\qw
}
\end{array}
\;.
\label{eq-n-two-to-n-three}
\eeq
Note that Eq.(\ref{eq-n-two-to-n-three})
reduces an $n^3$ controlled NOT
into a sequence of $n^2$ controlled NOTs.

\claim

For any real number $\theta$,

\grayeq{
\beq
\begin{array}{c}
\Qcircuit @C=1em @R=1em @!R{
&\ovalgate{\pi_1}\qwx[1]
&\qw
&\qw
\\
&\timesgate
&\gate{e^{i\theta\sigz}}
&\qw
}
\end{array}
=
\begin{array}{c}
\Qcircuit @C=1em @R=1em @!R{
&\gate{\pi_1}\qwx[1]
&\qw
&\gate{\pi_1}\qwx[1]
&\rstick{\vec{\bita}}\qw
\\
&\gate{e^{-2i\theta\sigz}}
&\gate{e^{i\theta\sigz}}
&\timesgate
&\rstick{\bitb}\qw
\gategroup{1}{2}{2}{2}{.7em}{.}
}
\end{array}
\;\;\;.
\eeq}
(Dotted box encloses wake.)
\proof
Consider Eq.(\ref{eq-perm-two-pi-contr-u})
with the following replacements:
$U_1\rarrow \sigx(\bitb)$,
$U_2\rarrow e^{i\theta \sigz(\bitb)}$,
$\pi_2\rarrow 1$.
Thus,

\beq
U_1 U_2 U_1^\dagger U_2^\dagger
\rarrow
\sigx(\bitb)
e^{i\theta \sigz(\bitb)}
\sigx(\bitb)
e^{-i\theta \sigz(\bitb)}
=e^{-2i\theta \sigz(\bitb)}
\;.
\eeq
\qed
\section{Exchanger}

We define the Exchanger
(a.k.a. Swapper or Exchange Operator
or Bit Transposition) by

\grayeq{
\beq
E(\bita, \bitb)
\ket{a, b}_{\bita\bitb}
=
\ket{b, a}_{\bita\bitb}
\;,
\eeq}
for all $a, b \in Bool$.
Therefore

\grayeq{
\beq
E(\bita, \bitb) = E(\bitb, \bita)
\;,
\eeq}
and

\grayeq{
\beq
E(\bita, \bitb)^2 = 1
\;.
\eeq}
Throughout QC Paulinesia,
we will represent Exchanger by

\grayeq{
\beq
E(\bita, \bitb)
=
\begin{array}{c}
\Qcircuit @C=1em @R=1em @!R{
&\uarrowgate\qwx[1]
&\qw
\\
&\darrowgate
&\qw
}
\end{array}
\;.
\eeq}

\claim

\grayeq{
\beq
E(\bita, \bitb)=
\cnot{\bitb}{\bita}
\cnot{\bita}{\bitb}
\cnot{\bitb}{\bita}
=
\begin{array}{c}
\Qcircuit @C=1em @R=1em @!R{
&\timesgate\qwx[1]
& \dotgate\qwx[1]
& \timesgate\qwx[1]
&\qw \\
&\dotgate
& \timesgate
& \dotgate
& \qw
}
\end{array}
\;.
\eeq}
\proof
\beqa
\cnot{\bitb}{\bita}
\cnot{\bita}{\bitb}
\cnot{\bitb}{\bita}
\ket{a,b}_{\bita\bitb}&=&
\cnot{\bitb}{\bita}
\cnot{\bita}{\bitb}
\ket{a \oplus b,b}\\
&=&
\cnot{\bitb}{\bita}
\ket{a \oplus b,a}\\
&=&
\ket{b,a}
\;.
\eeqa
\qed

\claim
If  $U$ and $V$
are $2\times2$ unitary matrices, then

\grayeq{
\beq
\begin{array}{c}
\Qcircuit @C=1em @R=1em @!R{
&\gate{U}
&\uarrowgate\qwx[1]
&\gate{V^\dagger}
&\qw
\\
&\gate{V}
&\darrowgate
&\gate{U^\dagger}
&\qw
}
\end{array}
=
\begin{array}{c}
\Qcircuit @C=1em @R=1em @!R{
&\uarrowgate\qwx[1]
&\qw
\\
&\darrowgate
&\qw
}
\end{array}
\;.
\label{eq-exc-two-u-invar}
\eeq}
\proof
Obvious.
\qed

\claim

\grayeq{
\beq
\begin{array}{c}
\Qcircuit @C=1em @R=1em @!R{
&\timesgate\qwx[1]
& \dotgate\qwx[1]
& \timesgate\qwx[1]
&\qw
\\
&\dotgate
& \timesgate
& \dotgate
&\qw
}
\end{array}
=
\begin{array}{c}
\Qcircuit @C=1em @R=1em @!R{
&\timesgate\qwx[1]
& \ogate\qwx[1]
& \timesgate\qwx[1]
&\qw
 \\
&\ogate
& \timesgate
& \ogate
&\qw
}
\end{array}
=
\begin{array}{c}
\Qcircuit @C=1em @R=1em @!R{
&\ogate
& \timesgate
& \ogate
&\qw
 \\
&\timesgate\qwx[-1]
& \ogate\qwx[-1]
& \timesgate\qwx[-1]
&\qw
}
\end{array}
=
\begin{array}{c}
\Qcircuit @C=1em @R=1em @!R{
&\dotgate
& \timesgate
& \dotgate
&\qw
 \\
&\timesgate\qwx[-1]
& \dotgate\qwx[-1]
& \timesgate\qwx[-1]
&\qw
}
\end{array}
\;.
\eeq
}
\proof
By virtue of Eq.(\ref{eq-exc-two-u-invar}),

\beq
\begin{array}{c}
\Qcircuit @C=1em @R=1em @!R{
&\timesgate\qwx[1]
& \dotgate\qwx[1]
& \timesgate\qwx[1]
&\qw
\\
&\dotgate
& \timesgate
& \dotgate
&\qw
}
\end{array}
=
\begin{array}{c}
\Qcircuit @C=1em @R=1em @!R{
&\gate{\sigx}
&\timesgate\qwx[1]
& \dotgate\qwx[1]
& \timesgate\qwx[1]
&\gate{\sigx}
&\qw
 \\
&\gate{\sigx}
&\dotgate
& \timesgate
& \dotgate
&\gate{\sigx}
&\qw
}
\end{array}
=
\begin{array}{c}
\Qcircuit @C=1em @R=1em @!R{
&\timesgate\qwx[1]
& \ogate\qwx[1]
& \timesgate\qwx[1]
&\qw
\\
&\ogate
& \timesgate
& \ogate
&\qw
}
\end{array}
\;.
\eeq
Likewise,

\beq
\begin{array}{c}
\Qcircuit @C=1em @R=1em @!R{
&\timesgate\qwx[1]
& \dotgate\qwx[1]
& \timesgate\qwx[1]
&\qw
\\
&\dotgate
& \timesgate
& \dotgate
&\qw
}
\end{array}
=
\begin{array}{c}
\Qcircuit @C=1em @R=1em @!R{
&\gate{H}
&\timesgate\qwx[1]
& \dotgate\qwx[1]
& \timesgate\qwx[1]
&\gate{H}
&\qw
 \\
&\gate{H}
&\dotgate
& \timesgate
& \dotgate
&\gate{H}
&\qw
}
\end{array}
=
\begin{array}{c}
\Qcircuit @C=1em @R=1em @!R{
&\dotgate
& \timesgate
& \dotgate
&\qw
 \\
&\timesgate\qwx[-1]
& \dotgate\qwx[-1]
& \timesgate\qwx[-1]
&\qw
}
\end{array}
\;.
\eeq
\qed

\claim

\grayeq{
\beq
E(\bita, \bitb)
=
[n(\bita)n(\bitb) + \nbar(\bita)\nbar(\bitb)]
+
\sigx(\bita)\sigx(\bitb)
[n(\bita)\nbar(\bitb) + \nbar(\bita)n(\bitb)]
\;.
\label{eq-exc-n-nbar}
\eeq}
\proof
Let RHS be the right hand side
of Eq.(\ref{eq-exc-n-nbar}).
For any $a,b\in Bool$,
if $a=b$, $RHS\ket{a,b}=\ket{a,b}$,
whereas when $a\neq b$,
$RHS\ket{a,b}=\ket{\overline{a},\overline{b}}$.
\qed

For any $x,z\in Bool$ and bit $\bita$, let
$\Lambda^{x,z}(\bita) =
\sigx^x(\bita)
\sigz^z(\bita)$. Note that
$[\Lambda^{x,z}]^\dagger
 = (-1)^{xz}\Lambda^{x,z}$
 and that
 $\Lambda^{00} = 1$,
 $\Lambda^{10} = \sigx$,
 $\Lambda^{11} = (-i)\sigy$,
 $\Lambda^{01} = \sigz$.
 As usual, let
 $\sigma_{w_1 w_2} =
 \sigma_{w_1}\otimes\sigma_{w_2}$
 for $w_1, w_2 \in \{X,Y,Z\}$.

\claim

\grayeq{
\beqa
E(\bita, \bitb)
&=&
\frac{1}{2}
\sum_{(x,z)\in Bool^2}
\Lambda^{xz}(\bita)
[\Lambda^{xz}(\bitb)]^\dagger\\
&=&
\frac{1}{2}
( 1
+ \sigma_{XX}
+ \sigma_{YY}
+ \sigma_{ZZ})
(\bita, \bitb)\\
&=&
\frac{1}{2}[
1 + \vec{\sigma}(\bita)
\cdot \vec{\sigma}(\bitb)]
\label{eq-exc-dot-prod}
\;.
\eeqa}
\proof
\beqa
\lefteqn{\frac{1}{2}
\sum_{x,z}
\Lambda^{xz}(\bita)(-1)^{xz}
\Lambda^{xz}(\bitb)
\ket{a,b}_{\bita, \bitb} =}\\
&=&
\frac{1}{2}
\sum_{x,z}
(-1)^{xz}
(\sigx^x\sigz^z \ket{a}_\bita)
(\sigx^x\sigz^z \ket{b}_\bitb)\\
&=&
\frac{1}{2}
\sum_{x,z}
(-1)^{(x+a+b)z}
\ket{a\oplus x, b \oplus x}\\
&=&
\frac{1}{2}
\sum_x 2 \delta^x_{a\oplus b}
\ket{a\oplus x, b \oplus x}\\
&=&
\ket{b,a}
\;.
\eeqa
\altproof
Replace the 3 CNOTs
in
$E(\bita, \bitb) =
\cnot{\bitb}{\bita}
\cnot{\bita}{\bitb}
\cnot{\bitb}{\bita}$
by
$\cnot{\bitb}{\bita}=
\frac{1}{2} \sum_{x,z} \sigx^x(\bita)
\sigz^z(\bitb)(-1)^{xz}$.
Details left to the reader.
\qed

We could
have predicted that
 $E(\bita, \bitb)$ would have the form
 Eq.(\ref{eq-exc-dot-prod})
 due to the invariance of Exchanger
 under identical rotations of both bits;
 that is, due to
 Eq.(\ref{eq-exc-two-u-invar})
 with
 $U=V=e^{i\vec{\theta}
 \cdot \vec{\sigma}}$,
 where $\vec{\theta}$ is
 an arbitrary
 3 dimensional real vector.

 \claim

 \grayeq{
\beq
\begin{array}{c}
\Qcircuit @C=1em @R=1em @!R{
&\uarrowgate\qwx[2]
&\qw
\\
&\qw
&\qw
\\
&\darrowgate
&\qw
}
\end{array}
=
\begin{array}{c}
\Qcircuit @C=1em @R=1em @!R{
&\uarrowgate\qwx[1]
&\qw
&\uarrowgate\qwx[1]
&\qw
\\
&\darrowgate
&\uarrowgate\qwx[1]
&\darrowgate
&\qw
\\
&\qw
&\darrowgate
&\qw
&\qw
}
\end{array}
\;.
\eeq}
\proof
Check that both sides map
$\bita\rarrow\bitc$,
$\bitb\rarrow\bitb$,
$\bitc\rarrow\bita$.
\qed
\section{Bell States}

Define the Bell state
$\ket{B^{00}}$ by

\grayeq{
\beq
\ket{B^{00}}=
\frac{1}{\sqrt{2}}(
\ket{00} + \ket{11} )
\;.
\eeq}

\claim

\grayeq{
\beq
\ket{B^{00}}_{\bita\bitb}=
\begin{array}{c}
\Qcircuit @C=1em @R=.5em @!R{
&\qw
&\dotgate\qwx[1]
&\gate{H}
&\gate{\ket{0}}
\\
&\qw
&\timesgate
&\qw
&\gate{\ket{0}}
}
\end{array}
\;.
\eeq}
\proof
\beqa
\cnot{\bita}{\bitb} H(\bita)
\ket{00}_{\bita\bitb}&=&
\sum_{a\in Bool}
\cnot{\bita}{\bitb}
\ket{a}_\bita
\bra{a}_\bita
H(\bita)
\ket{00}\\
&=&
\sum_a
\sigx^a(\bitb)
\ket{a,0}_{\bita\bitb}
(\frac{1}{\sqrt{2}})\\
&=&
\frac{1}{\sqrt{2}}
\sum_a \ket{a,a}
\;.
\eeqa
\qed

\claim

\grayeq{
\beq
\begin{array}{c}
\Qcircuit @C=1em @R=.5em @!R{
&\qw
&\multigate{1}{\ket{B^{00}}}
\\
&\gate{\sigx}
&\ghost{\ket{B^{00}}}
}
\end{array}
=
\begin{array}{c}
\Qcircuit @C=1em @R=.5em @!R{
&\gate{\sigx}
&\multigate{1}{\ket{B^{00}}}
\\
&\qw
&\ghost{\ket{B^{00}}}
}
\end{array}
\;,
\eeq}

\grayeq{
\beq
\begin{array}{c}
\Qcircuit @C=1em @R=.5em @!R{
&\qw
&\multigate{1}{\ket{B^{00}}}
\\
&\gate{\sigz}
&\ghost{\ket{B^{00}}}
}
\end{array}
=
\begin{array}{c}
\Qcircuit @C=1em @R=.5em @!R{
&\gate{\sigz}
&\multigate{1}{\ket{B^{00}}}
\\
&\qw
&\ghost{\ket{B^{00}}}
}
\end{array}
\;,
\eeq}

\grayeq{
\beq
\begin{array}{c}
\Qcircuit @C=1em @R=.5em @!R{
&\qw
&\multigate{1}{\ket{B^{00}}}
\\
&\gate{H}
&\ghost{\ket{B^{00}}}
}
\end{array}
=
\begin{array}{c}
\Qcircuit @C=1em @R=.5em @!R{
&\gate{H}
&\multigate{1}{\ket{B^{00}}}
\\
&\qw
&\ghost{\ket{B^{00}}}
}
\end{array}
\;.
\label{eq-move-h}
\eeq}
\proof
\beq
\sigx(\bitb)\sum_{a\in Bool}\ket{a,a}=
\sum_a \ket{a,\overline{a}}=
\sum_a \ket{\overline{a},a}=
\sigx(\bita)\sum_a\ket{a,a}
\;.
\eeq

\beq
\sigz(\bitb)\sum_{a\in Bool}\ket{a,a}=
\sum_a (-1)^a\ket{a,a}=
\sigz(\bita)\sum_a\ket{a,a}
\;.
\eeq

Eq.(\ref{eq-move-h}) follows from
the previous two equations and
the observation that $H = \frac{1}{\sqrt{2}}
(\sigx + \sigz)$.
\qed

Define the Bell states $\ket{B^{x,z}}$
and $\ket{B_{x,z}}$ for $x,z\in Bool$ by

\grayeq{
\beq
\ket{B_{x,z}}=
\begin{array}{c}
\Qcircuit @C=1em @R=.5em @!R{
&\qw
&\multigate{1}{\ket{B^{00}}}
\\
&\gate{\sigx^x\sigz^z}
&\ghost{\ket{B^{00}}}
}
\end{array}
\;,
\eeq}
and

\grayeq{
\beq
\ket{B^{x,z}}=
\begin{array}{c}
\Qcircuit @C=1em @R=.5em @!R{
&\gate{\sigx^x\sigz^z}
&\multigate{1}{\ket{B^{00}}}
\\
&\qw
&\ghost{\ket{B^{00}}}
}
\end{array}
\;.
\eeq}
Note that $\ket{B^{00}}=\ket{B_{00}}$. Since

\grayeq{
\beqa
\ket{B_{x,z}}&=&
\sigx^x(\bitb)\sigz^z(\bitb)
(\frac{1}{\sqrt{2}})
(\ket{00} + \ket{11})_{\bita\bitb}\\
&=&
\frac{1}{\sqrt{2}}
(\ket{0x} + (-1)^z\ket{1\overline{x}})
\;,
\eeqa}
it follows that

\grayeq{
\beq
\begin{array}{rrr}
\ket{B_{00}} =
&1\ket{B_{00}}=
&\frac{1}{\sqrt{2}}
(\ket{00} + \ket{11})
\\
\ket{B_{10}} =
&\sigx(\bitb)\ket{B_{00}}=
&\frac{1}{\sqrt{2}}
(\ket{01} + \ket{10})
\\
\ket{B_{11}} =
&(-i)\sigy(\bitb)\ket{B_{00}}=
&\frac{1}{\sqrt{2}}
(\ket{01} - \ket{10})
\\
\ket{B_{01}} =
&\sigz(\bitb)\ket{B_{00}}=
&\frac{1}{\sqrt{2}}
(\ket{00} - \ket{11})
\end{array}
\;.
\eeq}

\claim

For any $x,z\in Bool$,

\grayeq{
\beqa
\ket{B_{x,z}}_{\bita\bitb}&=&
E(\bita, \bitb)
\ket{B^{x,z}}_{\bita\bitb}
\label{eq-e-raises-xz}
\\
&=&
(-1)^{xz}
\ket{B^{x,z}}_{\bita\bitb}
\;.
\label{eq-bell-eva}
\eeqa}
Thus,
$\ket{B^{x,z}}$ and $\ket{B_{x,z}}$
are both eigenfunctions of $E$ with
eigenvalue $(-1)^{xz}$.
\proof
Eq.(\ref{eq-e-raises-xz}) is
obvious. Eq.(\ref{eq-bell-eva})
follows from

\beqa
\begin{array}{c}
\Qcircuit @C=1em @R=.5em @!R{
&\qw
&\multigate{1}{\ket{B^{00}}}
\\
&\gate{\sigx^x\sigz^z}
&\ghost{\ket{B^{00}}}
}
\end{array}
&=&
\begin{array}{c}
\Qcircuit @C=1em @R=.5em @!R{
&\gate{\sigz^z}
&\multigate{1}{\ket{B^{00}}}
\\
&\gate{\sigx^x}
&\ghost{\ket{B^{00}}}
}
\end{array}
\\
&=&
\begin{array}{c}
\Qcircuit @C=1em @R=.5em @!R{
&\gate{\sigz^z\sigx^x}
&\multigate{1}{\ket{B^{00}}}
\\
&\qw
&\ghost{\ket{B^{00}}}
}
\end{array}\\
&=&(-1)^{xz}
\begin{array}{c}
\Qcircuit @C=1em @R=.5em @!R{
&\gate{\sigx^x\sigz^z}
&\multigate{1}{\ket{B^{00}}}
\\
&\qw
&\ghost{\ket{B^{00}}}
}
\end{array}
\;.
\eeqa
\qed

\claim

For any $x,z\in Bool$,

\grayeq{
\beq
\ket{B_{x,z}}=
\begin{array}{c}
\Qcircuit @C=1em @R=.5em @!R{
&\dotgate\qwx[1]
&\gate{H}
&\gate{\ket{z}}
\\
&\timesgate
&\qw
&\gate{\ket{x}}
}
\end{array}
\;,
\eeq}

\grayeq{
\beq
\ket{B^{x,z}}=
\begin{array}{c}
\Qcircuit @C=1em @R=.5em @!R{
&\timesgate\qwx[1]
&\qw
&\gate{\ket{x}}
\\
&\dotgate
&\gate{H}
&\gate{\ket{z}}
}
\end{array}
\;.
\eeq}
\proof
\beqa
\begin{array}{c}
\Qcircuit @C=1em @R=.5em @!R{
&\qw
&\multigate{1}{\ket{B^{00}}}
\\
&\gate{\sigx^x\sigz^z}
&\ghost{\ket{B^{00}}}
}
\end{array}
&=&
\begin{array}{c}
\Qcircuit @C=1em @R=.5em @!R{
&\gate{\sigz^z}
&\multigate{1}{\ket{B^{00}}}
\\
&\gate{\sigx^x}
&\ghost{\ket{B^{00}}}
}
\end{array}\\
&=&
\begin{array}{c}
\Qcircuit @C=1em @R=.5em @!R{
&\gate{\sigz^z}
&\dotgate\qwx[1]
&\gate{H}
&\gate{\ket{0}}
\\
&\gate{\sigx^x}
&\timesgate
&\qw
&\gate{\ket{0}}
}
\end{array}\\
&=&
\begin{array}{c}
\Qcircuit @C=1em @R=.5em @!R{
&\dotgate\qwx[1]
&\gate{H}
&\gate{\ket{z}}
\\
&\timesgate
&\qw
&\gate{\ket{x}}
}
\end{array}
\;.
\eeqa
\qed

\claim (Orthonormality)

\grayeq{
\beq
\av{B_{xz}|B_{x'z'}} = \delta^{x',z'}_{x,z}
\;
\eeq}
for any $x,z,x'z'\in Bool$, and

\grayeq{
\beq
\sum_{(x,z)\in Bool^2}\ket{B_{xz}}\bra{B_{xz}} = 1
\;.
\eeq}
\proof
\beq
\begin{array}{c}
\Qcircuit @C=1em @R=.5em @!R{
&\freegate{\bra{z'}}
&\gate{H}
&\dotgate\qwx[1]
&\dotgate\qwx[1]
&\gate{H}
&\gate{\ket{z}}
\\
&\freegate{\bra{x'}}
&\qw
&\timesgate
&\timesgate
&\qw
&\gate{\ket{x}}
}
\end{array}
=
\delta^{x',z'}_{x,z}
\;.
\eeq

\beq
\sum_{x,z}
\begin{array}{c}
\Qcircuit @C=1em @R=.5em @!R{
&\qw
&\dotgate\qwx[1]
&\gate{H}
&\gate{\ket{z}\bra{z}}
&\gate{H}
&\dotgate\qwx[1]
&\qw
\\
&\qw
&\timesgate
&\qw
&\gate{\ket{x}\bra{x}}
&\qw
&\timesgate
&\qw
}
\end{array}
=1
\;.
\eeq
\qed

\claim

\grayeq{
For all $a,b,x,z\in Bool$,
if $P(a,b|x,z)=|\av{a,b|B_{x,z}}|^2$,
then the marginals $P(a|x,z)$
and $P(b|x,z)$ are both identically equal to
$\frac{1}{2}$.}
\proof
\beqa
\sum_a P(a,b|x,z) &=&
\sum_a
\begin{array}{c}
\Qcircuit @C=1em @R=.5em @!R{
\freegate{\bra{z}}
&\gate{H}
&\dotgate\qwx[1]
&\gate{\ket{a}\bra{a}}
&\dotgate\qwx[1]
&\gate{H}
&\gate{\ket{z}}
\\
\freegate{\bra{x}}
&\qw
&\timesgate
&\gate{\ket{b}\bra{b}}
&\timesgate
&\qw
&\gate{\ket{x}}
}
\end{array}
\\
&=&
\sum_a
\left|
\frac{(-1)^{za}}{\sqrt{2}}
\bra{x} \sigx^a \ket{b}
\right|^2\\
&=&
\frac{1}{2}
\sum_a \left|
\av{x|a\oplus b}
\right|^2
\\
&=&
\frac{1}{2}\sum_a \delta_a^{x\oplus b}
= \frac{1}{2}
\;.
\eeqa
\qed
\section{GHZ}

The GHZ state is defined by

\grayeq{
\beq
\ket{GHZ} =
\frac{1}{\sqrt{2}}
(\ket{000} + \ket{111})
\;.
\eeq}

\claim

\grayeq{
\beq
\ket{GHZ} =
\begin{array}{c}
\Qcircuit @C=1em @R=.5em @!R{
&\timesgate\qwx[2]
&\qw
&\qw
&\gate{\ket{0}}
\\
&\qw
&\timesgate\qwx[1]
&\qw
&\gate{\ket{0}}
\\
&\dotgate
&\dotgate
&\gate{H}
&\gate{\ket{0}}
}
\end{array}
\;.
\label{eq-ghz-ckt}
\eeq}
\proof
Let RHS denote right
hand side of Eq.(\ref{eq-ghz-ckt}).

\beqa
RHS&=&
\cnot{\bitc}{\bita}
\cnot{\bitc}{\bitb}
H(\bitc)
\ket{000}_{\bita\bitb\bitc}
\\
&=&
\cnot{\bitc}{\bita}
\cnot{\bitc}{\bitb}
\frac{1}{\sqrt{2}}\sum_{a\in Bool}\ket{0,0,a}
\\
&=&
\frac{1}{\sqrt{2}}\sum_{a\in Bool}\ket{a,a,a}=\ket{GHZ}
\;.
\eeqa
\qed

\claim

\grayeq{
\beq
\sigma_{XYY}\ket{GHZ}=
\sigma_{YXY}\ket{GHZ}=
\sigma_{YYX}\ket{GHZ}=-\ket{GHZ}
\;.
\label{eq-ghz-xyy}
\eeq}
Hence,

\grayeq{
\beq
\sigma_{XYY}\sigma_{YXY}
\sigma_{YYX}\ket{GHZ}=-\ket{GHZ}
\;.
\label{eq-ghz-3xyy}
\eeq}
However,

\grayeq{
\beq
\sigma_{XXX}\ket{GHZ}=+\ket{GHZ}
\;.
\label{eq-ghz-xxx}
\eeq}
\proof
For any $a\in Bool$,
$\sigy\ket{a}=i(-1)^a\ket{\overline{a}}$ and
$\sigx\ket{a}=\ket{\overline{a}}$ so

\beqa
\sigma_{XYY}\ket{GHZ}&=&
\sigx\otimes\sigy\otimes\sigy
\frac{1}{\sqrt{2}}\sum_{a\in Bool}\ket{a,a,a}
\\
&=&
(-1)\frac{1}{\sqrt{2}}
\sum_a\ket{\overline{a},\overline{a},\overline{a}}\\
&=& -\ket{GHZ}
\;.
\eeqa
This establishes Eq.(\ref{eq-ghz-xyy}).
Eq.(\ref{eq-ghz-3xyy}) follows from
Eq.(\ref{eq-ghz-xyy}).
Eq.(\ref{eq-ghz-xxx}) can be proven in the
same way as Eq.(\ref{eq-ghz-xyy}).
\qed
\section{One and Two Qubit Projective Measurements}

\claim (Conversion: 1 qubit internal measurement $\rarrow$
1 qubit final measurement)

For any $j\in Bool$,

\grayeq{
\beq
\begin{array}{c}
\Qcircuit @C=1em @R=1em @!R{
&
&
\\
&\gate{\ket{j}\bra{j}}
&\qw
}
\end{array}
=
\begin{array}{c}
\Qcircuit @C=1em @R=.5em @!R{
\freegate{\bra{j}}
&\timesgate\qwx[1]
&\gate{\ket{0}}
\\
&\dotgate
&\qw
}
\end{array}
\;.
\label{eq-1qbit-proj}
\eeq}
\proof
Let LHS and RHS stand for the left and
right hand sides of Eq.(\ref{eq-1qbit-proj}).
For any $b\in Bool$,

\beq
LHS\ket{b}_\bitb =
\ket{b}_\bitb \delta_j^b
\;.
\eeq

\beqa
RHS\ket{b}_\bitb &=&
\bra{j}_\bita
\cnot{\bitb}{\bita}
\ket{0,b}_{\bita\bitb}\\
&=&
\bra{j}_\bita \ket{b,b}_{\bita\bitb}\\
&=&
\ket{b}_\bitb \delta_j^b
\;.
\eeqa
\qed

One qubit operations (such as
internal or final one qubit measurements
or one qubit rotations)
are ``cheap" compared with two qubit operations
such as CNOTs and two qubit measurements (either
internal or final). This is because two
qubit operations are slower and
they require two qubits to interact,
which  opens the door for noise from
the environment to creep in.
So in this section we
will pay attention only to
the number of two qubit operations.
Let {\bf bibit} stand for two bits.
Next we will show what I like to
call the ``one to
two" conversion rules. Namely, given a single
CNOT, one can always convert it to two
bibit operations. Likewise,
given a single bibit operation, one can always
convert it to two CNOTs.

As usual in this document,
for $j\in Bool$, we define
$\pizz{j}(\bita, \bitb)=
\pi[\sigzz(\bita, \bitb)=(-1)^j]$; i.e,
$\pizz{j}$ is
the projection operator onto the
2 qubit subspace with $(-1)^j$
as eigenvalue for $\sigz\otimes\sigz$.

\claim (Conversion: 1 bibit
measurement $\rarrow$
2 CNOTs)

For any $j\in Bool$,

\grayeq{
\beq
\begin{array}{c}
\Qcircuit @C=1em @R=1.5em @!R{
&\multigate{1}{\pizz{j}}
&\qw
\\
&\ghost{\pizz{j}}
&\qw
}
\end{array}
=
\begin{array}{c}
\Qcircuit @C=1em @R=.5em @!R{
&\dotgate\qwx[1]
&\qw
&\dotgate\qwx[1]
&\qw
\\
&\timesgate
&\gate{\ket{j}\bra{j}}
&\timesgate
&\qw
}
\end{array}
\;.
\label{eq-1meas-to-2cnots-2bit}
\eeq}
\proof
Let RHS stand for the right
hand side of Eq.(\ref{eq-1meas-to-2cnots-2bit}).
For any $a,b\in Bool$,

\beqa
RHS\ket{a,b}_{\bita\bitb}&=&
\cnot{\bita}{\bitb}
\ket{j}_\bitb
\bra{j}_\bitb
\cnot{\bita}{\bitb}
\ket{a,b}_{\bita\bitb}\\
&=&
\cnot{\bita}{\bitb}
\ket{j}_\bitb
\delta^j_{a\oplus b}
\ket{a}_\bita\\
&=&
\delta^j_{a\oplus b}
\ket{a,b}_{\bita\bitb}\\
&=&
\pizz{j}\ket{a,b}_{\bita\bitb}
\;.
\eeqa
\qed

\claim (Conversion: 1 bibit measurement $\rarrow$
1 CNOT. Special case of
Eq.(\ref{eq-1meas-to-2cnots-2bit}).)

For any $j,k\in Bool$,

\grayeq{
\beq
\begin{array}{c}
\Qcircuit @C=1em @R=.5em @!R{
\freegate{\bra{k}}
&\multigate{1}{\pizz{j}}
&\qw
\\
&\ghost{\pizz{j}}
&\qw
}
\end{array}
=
\begin{array}{c}
\Qcircuit @C=1em @R=.5em @!R{
\freegate{\bra{k}}
&\qw
&\qw
&\dotgate\qwx[1]
&\qw
\\
&\gate{\sigx^k}
&\gate{\ket{j}\bra{j}}
&\timesgate
&\qw
}
\end{array}
\;.
\label{eq-1meas-to-1cnot}
\eeq}
\proof
Follows immediately from
Eq.(\ref{eq-1meas-to-2cnots-2bit}).
\qed

\claim (Another Conversion of: 1 bibit
measurement $\rarrow$
2 CNOTs)

For any $j\in Bool$,

\grayeq{
\beq
\begin{array}{c}
\Qcircuit @C=1em @R=1.5em @!R{
&
&
\\
&\multigate{1}{\pizz{j}}
&\qw
\\
&\ghost{\pizz{j}}
&\qw
}
\end{array}
=
\begin{array}{c}
\Qcircuit @C=1em @R=.5em @!R{
\freegate{\bra{j}}
&\timesgate\qwx[1]
&\timesgate\qwx[2]
&\gate{\ket{0}}
\\
&\dotgate
&\qw
&\qw
\\
&\qw
&\dotgate
&\qw
}
\end{array}
\;.
\label{eq-1meas-to-2cnots-3bit}
\eeq}
\proof
Let LHS and RHS denote the left and
right hand sides of
Eq.(\ref{eq-1meas-to-2cnots-3bit}).

\beqa
LHS &=&
\begin{array}{c}
\Qcircuit @C=1em @R=.5em @!R{
&
&
&
&
\\
&\dotgate\qwx[1]
&\qw
&\dotgate\qwx[1]
&\qw
\\
&\timesgate
&\gate{\ket{j}\bra{j}}
&\timesgate
&\qw
}
\end{array}\\
&=&
\begin{array}{c}
\Qcircuit @C=1em @R=.5em @!R{
\freegate{\bra{j}}
&\qw
&\timesgate\qwx[2]
&\qw
&\gate{\ket{0}}
\\
&\dotgate\qwx[1]
&\qw
&\dotgate\qwx[1]
&\qw
\\
&\timesgate
&\dotgate
&\timesgate
&\qw
}
\end{array}\\
&=&
\begin{array}{c}
\Qcircuit @C=1em @R=.5em @!R{
\freegate{\bra{j}}
&\timesgate\qwx[1]
&\timesgate\qwx[2]
&\qw
&\qw
&\gate{\ket{0}}
\\
&\dotgate
&\qw
&\dotgate\qwx[1]
&\dotgate\qwx[1]
&\qw
\\
&\qw
&\dotgate
&\timesgate
&\timesgate
&\qw
}
\end{array}\\
&=&RHS
\;.
\eeqa
\qed

\claim (Conversion: 1 CNOT $\rarrow$
2 bibit measurements)

For any $k, j_1, j_2\in Bool$,

\grayeq{
\beq
\begin{array}{c}
\Qcircuit @C=1em @R=2em @!R{
&\dotgate\qwx[2]
&\qw
\\
&
&
\\
&\timesgate
&\qw
}
\end{array}
=
(-1)^{(k+j_1)j_2}2\sqrt{2}
\begin{array}{c}
\Qcircuit @C=.5em @R=.5em @!R{
&\gate{\sigz^{j_2}}
&\qw
&\qw
&\qw
&\multigate{1}{\pizz{j_1}}
&\qw
&\qw
\\
\freegate{\bra{k}}
&\qw
&\gate{H}
&\multigate{1}{\pizz{j_2}}
&\gate{H}
&\ghost{\pizz{j_1}}
&\gate{H}
&\gate{\ket{0}}
\\
&\gate{\sigx^{k+j_1}}
&\gate{H}
&\ghost{\pizz{j_2}}
&\gate{H}
&\qw
&\qw
&\qw
}
\end{array}
\;.
\label{eq-1cnot-to-2meas}
\eeq}
\proof
Define $T$ by

\beq
T=
\begin{array}{c}
\Qcircuit @C=.5em @R=.5em @!R{
&\qw
&\qw
&\qw
&\multigate{1}{\pizz{j_1}}
&\qw
&\qw
\\
\freegate{\bra{k}}
&\gate{H}
&\multigate{1}{\pizz{j_2}}
&\gate{H}
&\ghost{\pizz{j_1}}
&\gate{H}
&\gate{\ket{0}}
\\
&\gate{H}
&\ghost{\pizz{j_2}}
&\gate{H}
&\qw
&\qw
&\qw
}
\end{array}
\;.
\eeq
Then

\beq
T=
\underbrace{
\begin{array}{c}
\Qcircuit @C=.5em @R=.5em @!R{
&\qw
&\qw
&\qw
\\
\freegate{\bra{k}}
&\gate{H}
&\timesgate\qwx[1]
&\gate{\ket{j_2}}
\\
&\gate{H}
&\dotgate
&\qw
}
\end{array}
}_{T_1}
\;
\underbrace{
\begin{array}{c}
\Qcircuit @C=.5em @R=.5em @!R{
&\qw
&\qw
&\dotgate\qwx[1]
&\qw
\\
\freegate{\bra{j_2}}
&\timesgate\qwx[1]
&\gate{H}
&\timesgate
&\gate{\ket{j_1}}
\\
&\dotgate
&\gate{H}
&\qw
&\qw
}
\end{array}
}_{T_2}
\;
\underbrace{
\begin{array}{c}
\Qcircuit @C=.5em @R=.5em @!R{
&\dotgate\qwx[1]
&\qw
&\qw
&\qw
\\
\freegate{\bra{j_1}}
&\timesgate
&\gate{H}
&\gate{\ket{0}}
\\
&\qw
&\qw
&\qw
}
\end{array}
}_{T_3}
\;\;,
\eeq

\beq
T_1 =
\frac{(-1)^{k j_2}}{\sqrt{2}}
H(\bitc)
\sigz^k(\bitc)
\;,
\eeq

\beq
T_3 = \frac{1}{\sqrt{2}}
\;,
\eeq

\beqa
T_2 &=&
\begin{array}{c}
\Qcircuit @C=1em @R=.5em @!R{
&\qw
&\qw
&\dotgate\qwx[1]
&\qw
\\
\freegate{\bra{j_2}}
&\gate{H}
&\dotgate\qwx[1]
&\timesgate
&\gate{\ket{j_1}}
\\
&\gate{H}
&\timesgate
&\qw
&\qw
}
\end{array}\\
&=&
\begin{array}{c}
\Qcircuit @C=1em @R=.5em @!R{
&\qw
&\dotgate\qwx[2]
&\dotgate\qwx[1]
&\qw
&\qw
\\
\freegate{\bra{j_2}}
&\gate{H}
&\qw
&\timesgate
&\dotgate\qwx[1]
&\gate{\ket{j_1}}
\\
&\gate{H}
&\timesgate
&\qw
&\timesgate
&\qw
}
\end{array}\\
&=&
\frac{(-1)^{j_1 j_2}}{\sqrt{2}}
\begin{array}{c}
\Qcircuit @C=1em @R=.5em @!R{
&\qw
&\dotgate\qwx[2]
&\gate{\sigz^{j_2}}
\\
&
&
&
\\
&\gate{H}
&\timesgate
&\gate{\sigx^{j_1}}
}
\end{array}
\;.
\eeqa
Putting all this together,

\beqa
T &=& T_1 T_2 T_3\\
&=&
\frac{(-1)^{(k+j_1) j_2}}{2\sqrt{2}}
\begin{array}{c}
\Qcircuit @C=1em @R=.5em @!R{
&\qw
&\dotgate\qwx[1]
&\gate{\sigz^{j_2}}
\\
&\gate{H\sigz^k H}
&\timesgate
&\gate{\sigx^{j_1}}
}
\end{array}\\
&=&
\frac{(-1)^{(k+j_1) j_2}}{2\sqrt{2}}
\begin{array}{c}
\Qcircuit @C=1em @R=.5em @!R{
&\gate{\sigz^{j_2}}
&\dotgate\qwx[1]
&\qw
\\
&\gate{\sigx^{k+j_1}}
&\timesgate
&\qw
}
\end{array}
\;.
\eeqa
\altproof
Define operator $S$ such that for all
$a,c\in Bool$,
\beq
S\ket{a,c}_{\bita\bitc}=
\bra{k}_\bitb H(\bitb)
\pizz{j_2}(\bitb,\bitc)
H(\bitb)
\pizz{j_1}(\bita, \bitb)
H(\bitb)
\ket{a,0,c}_{\bita\bitb\bitc}
\;.
\label{eq-alg-t}
\eeq
In Eq.(\ref{eq-alg-t}),
insert a partition of unity
$\sum_{(a_1, b_1, c_1)\in Bool^3}
\ket{a_1,b_1, c_1}
\bra{a_1,b_1, c_1}$
before the first bibit
measurement and
another
$\sum_{(a_2, b_2, c_2)\in Bool^3}
\ket{a_2,b_2, c_2}
\bra{a_2,b_2, c_2}$
before the second.
Then use the fact that
for $a,b,j\in Bool$,
$\pizz{j}\ket{a,b} = \delta^j_{a\oplus b}\ket{a,b}$.
Details left to the reader.
\qed

\claim (Conversion: 1 CNOT $\rarrow$ 1
bibit measurement.
Special case of
Eq.(\ref{eq-1cnot-to-2meas}).)

For any $j,k\in Bool$,

\grayeq{
\beq
\begin{array}{c}
\Qcircuit @C=1em @R=.5em @!R{
&\dotgate\qwx[1]
&\qw
\\
&\timesgate
&\gate{\ket{j}}
}
\end{array}
=
(-1)^{jk}\sqrt{2}
\begin{array}{c}
\Qcircuit @C=1em @R=.5em @!R{
&\gate{\sigz^j}
&\multigate{1}{\pizz{j}}
&\qw
&\qw
\\
&\qw
&\ghost{\pizz{j}}
&\gate{H}
&\gate{\ket{k}}
}
\end{array}
\;.
\label{eq-1cnot-1meas}
\eeq}
\proof
Let LHS and RHS stand for the left and
right hand sides of Eq.(\ref{eq-1cnot-1meas}).
Then

\beq
RHS=
(-1)^{jk}\sqrt{2}
\begin{array}{c}
\Qcircuit @C=1em @R=.5em @!R{
&\gate{\sigz^j}
&\dotgate\qwx[1]
&\qw
&\dotgate\qwx[1]
&\qw
&\qw
\\
&\qw
&\timesgate
&\gate{\ket{j}\bra{j}}
&\timesgate
&\gate{H}
&\gate{\ket{k}}
}
\end{array}
=
LHS
\;.
\eeq
\qed
\section{Two Qubit Exchange Scattering}
Throughout this section, $\ket{\psi}$
will denote an arbitrary one qubit state.

\claim (Exchange scattering via Exchanger)

For any $z\in Bool$,

\grayeq{
\beq
\sqrt{2}
\begin{array}{c}
\Qcircuit @C=1em @R=.5em @!R{
\freegate{\bra{z}}
&\gate{H}
&\uarrowgate\qwx[1]
&\gate{\ket{\psi}}
\\
&\qw
&\darrowgate
&\gate{\ket{0}}
}
\end{array}
=
\begin{array}{c}
\Qcircuit @C=1em @R=.5em @!R{
&
\\
& \gate{\ket{\psi}}
}
\end{array}
\;.
\label{eq-ex-scat-via-exch}
\eeq}
\proof
Let LHS and RHS stand for the left and
right hand sides of
Eq.(\ref{eq-ex-scat-via-exch}).

\beq
LHS =
\sqrt{2}
\begin{array}{c}
\Qcircuit @C=1em @R=.5em @!R{
\freegate{\bra{z}}
&\gate{H}
&\gate{\ket{0}}
\\
&\qw
&\gate{\ket{\psi}}
}
\end{array}
= RHS
\;.
\eeq
\qed

\claim (Exchange scattering via CNOT)

For any $z\in Bool$,

\grayeq{
\beq
\sqrt{2}
\begin{array}{c}
\Qcircuit @C=1em @R=.5em @!R{
\freegate{\bra{z}}
&\qw
&\gate{H}
&\dotgate\qwx[1]
&\gate{\ket{\psi}}
\\
&\gate{\sigz^z}
&\qw
&\timesgate
&\gate{\ket{0}}
}
\end{array}
=
\begin{array}{c}
\Qcircuit @C=1em @R=.5em @!R{
&
\\
& \gate{\ket{\psi}}
}
\end{array}
\;.
\label{eq-ex-scat-via-cnot1}
\eeq}
\proof
Let LHS and RHS stand for the left and
right hand sides of
Eq.(\ref{eq-ex-scat-via-cnot1}).

\beqa
LHS &=&
\sqrt{2}
\begin{array}{c}
\Qcircuit @C=1em @R=.5em @!R{
\freegate{\bra{z}}
&\dotgate\qwx[1]
&\gate{H}
&\dotgate\qwx[1]
&\uarrowgate\qwx[1]
&\gate{\ket{0}}
\\
&\dotgate
&\qw
&\timesgate
&\darrowgate
&\gate{\ket{\psi}}
}
\end{array}\\
&=&
\sqrt{2}
\begin{array}{c}
\Qcircuit @C=1em @R=.5em @!R{
\freegate{\bra{z}}
&\gate{H}
&\timesgate\qwx[1]
&\dotgate\qwx[1]
&\dotgate\qwx[1]
&\timesgate\qwx[1]
&\dotgate\qwx[1]
&\gate{\ket{0}}
\\
&\qw
&\dotgate
&\timesgate
&\timesgate
&\dotgate
&\timesgate
&\gate{\ket{\psi}}
}
\end{array}\\
&=&
\sqrt{2}
\begin{array}{c}
\Qcircuit @C=1em @R=.5em @!R{
\freegate{\bra{z}}
&\gate{H}
&\gate{\ket{0}}
\\
&\qw
&\gate{\ket{\psi}}
}
\end{array}\\
&=& RHS
\;.
\eeqa
\altproof
For any $a\in Bool$:

\beq
\sqrt{2}
\begin{array}{c}
\Qcircuit @C=1em @R=.5em @!R{
\freegate{\bra{0}H}
&\dotgate\qwx[1]
&\gate{\ket{a}}
\\
&\timesgate
&\gate{\ket{0}}
}
\end{array}
=
\sigx^a(\bitb)\ket{0}_\bitb
= \ket{a}_\bitb
\;.
\eeq
Thus, for an arbitrary state $\ket{\psi}$,

\beq
\sqrt{2}
\begin{array}{c}
\Qcircuit @C=1em @R=.5em @!R{
\freegate{\bra{0}H}
&\dotgate\qwx[1]
&\gate{\ket{\psi}}
\\
&\timesgate
&\gate{\ket{0}}
}
\end{array}
=
\begin{array}{c}
\Qcircuit @C=1em @R=.5em @!R{
&
\\
&\gate{\ket{\psi}}
}
\end{array}
\;.
\label{eq-ex-scat-lite}
\eeq
In Eq.(\ref{eq-ex-scat-lite}),
if we multiply ket $\ket{\psi}$
by a pre-processing and a post-processing $\sigz^z$,
then we obtain

\beq
\sqrt{2}
\begin{array}{c}
\Qcircuit @C=1em @R=.5em @!R{
\freegate{\bra{0}H}
&\qw
&\dotgate\qwx[1]
&\gate{\sigz^z\ket{\psi}}
\\
&\gate{\sigz^z}
&\timesgate
&\gate{\ket{0}}
}
\end{array}
=
\begin{array}{c}
\Qcircuit @C=1em @R=.5em @!R{
&
\\
&\gate{(\sigz^z)^2\ket{\psi}}
}
\end{array}
\;,
\eeq
which easily yields

\beq
\sqrt{2}
\begin{array}{c}
\Qcircuit @C=1em @R=.5em @!R{
\freegate{\bra{z}}
&\gate{H}
&\dotgate\qwx[1]
&\gate{\ket{\psi}}
\\
&\gate{\sigz^z}
&\timesgate
&\gate{\ket{0}}
}
\end{array}
=
\begin{array}{c}
\Qcircuit @C=1em @R=.5em @!R{
&
\\
&\gate{\ket{\psi}}
}
\end{array}
\;.
\eeq
\qed

\claim (Another example of exchange scattering via CNOT)

For any $x\in Bool$,

\grayeq{
\beq
\sqrt{2}
\begin{array}{c}
\Qcircuit @C=1em @R=.5em @!R{
\freegate{\bra{x}}
&\qw
&\timesgate\qwx[1]
&\qw
&\gate{\ket{\psi}}
\\
&\gate{\sigx^x}
&\dotgate
&\gate{H}
&\gate{\ket{0}}
}
\end{array}
=
\begin{array}{c}
\Qcircuit @C=1em @R=.5em @!R{
&
\\
& \gate{\ket{\psi}}
}
\end{array}
\;.
\label{eq-ex-scat-via-cnot2}
\eeq}
\proof
In Eq.(\ref{eq-ex-scat-via-cnot1}),
if we replace $z$ by $x$ and
multiply the ket $\ket{\psi}$
by a pre-processing and a
post-processing $H$, then we obtain:

\beq
\sqrt{2}
\begin{array}{c}
\Qcircuit @C=1em @R=.5em @!R{
\freegate{\bra{x}}
&\qw
&\dotgate\qwx[1]
&\gate{H}
&\dotgate\qwx[1]
&\gate{H\ket{\psi}}
\\
&\gate{H}
&\dotgate
&\qw
&\timesgate
&\gate{\ket{0}}
}
\end{array}
=
\begin{array}{c}
\Qcircuit @C=1em @R=.5em @!R{
&
\\
& \gate{H^2\ket{\psi}}
}
\end{array}
\;.
\eeq
The last identity simplifies to

\beq
\sqrt{2}
\begin{array}{c}
\Qcircuit @C=1em @R=.5em @!R{
\freegate{\bra{x}}
&\dotgate\qwx[1]
&\timesgate\qwx[1]
&\qw
&\gate{\ket{\psi}}
\\
&\timesgate
&\dotgate
&\gate{H}
&\gate{\ket{0}}
}
\end{array}
=
\begin{array}{c}
\Qcircuit @C=1em @R=.5em @!R{
&
\\
& \gate{\ket{\psi}}
}
\end{array}
\;,
\eeq
which is the same as the claim that we
set out to prove.
\qed

\claim (Exchange scattering via a 2 qubit
projective measurement)

For any $j,k\in Bool$,

\grayeq{
\beq
2
\begin{array}{c}
\Qcircuit @C=1em @R=.5em @!R{
\freegate{\bra{j}}
&\gate{H}
&\multigate{1}{\pizz{k}}
&\qw
&\gate{\ket{\psi}}
\\
&\gate{\sigz^j\sigx^k}
&\ghost{\pizz{k}}
&\gate{H}
&\gate{\ket{0}}
}
\end{array}
=
\begin{array}{c}
\Qcircuit @C=1em @R=.5em @!R{
&
\\
& \gate{\ket{\psi}}
}
\end{array}
\;.
\label{eq-ex-scat-via-proj}
\eeq}
\proof
\beqa
\lefteqn{
\begin{array}{c}
\Qcircuit @C=1em @R=.5em @!R{
\freegate{\bra{j}}
&\gate{H}
&\multigate{1}{\pizz{k}}
&\qw
&\gate{\ket{\psi}}
\\
&\qw
&\ghost{\pizz{k}}
&\gate{H}
&\gate{\ket{0}}
}
\end{array}
=}
\label{eq-scat-one-box}
\\
&=&
\begin{array}{c}
\Qcircuit @C=1em @R=.5em @!R{
\freegate{\bra{j}}
&\gate{H}
&\timesgate\qwx[1]
&\gate{\ket{k}}
&\freegate{\bra{k}}
&\timesgate\qwx[1]
&\qw
&\gate{\ket{\psi}}
\\
&\qw
&\dotgate
&\qw
&\qw
&\dotgate
&\gate{H}
&\gate{\ket{0}}
\gategroup{1}{1}{2}{4}{.7em}{.}
\gategroup{1}{5}{2}{8}{.7em}{.}
}
\end{array}
\label{eq-scat-two-boxes}
\\
&=&
\left[
\frac{(-1)^{jk}}{\sqrt{2}}
\sigz^j(\bitb)
\right]
\left[
\frac{\sigx^k(\bitb)}{\sqrt{2}}\ket{\psi}_\bitb
\right]
\label{eq-scat-no-boxes}
\\
&=&
\sigx^k(\bitb)\sigz^j(\bitb)
\ket{\psi}_\bitb/2
\;.
\eeqa
To go from Eq.(\ref{eq-scat-one-box})
to Eq.(\ref{eq-scat-two-boxes}), we
expressed the two qubit projective measurement
in terms of 2 CNOTs, as described in
the section entitled One and Two Qubit
Projective Measurements.
To go from Eq.(\ref{eq-scat-two-boxes})
to Eq.(\ref{eq-scat-no-boxes}), we used identity
Eq.(\ref{eq-ex-scat-via-cnot2})
to reduce the second dotted
box of Eq.(\ref{eq-scat-two-boxes}).
\altproof
For any $a\in Bool$,
\beqa
\lefteqn{
\begin{array}{c}
\Qcircuit @C=1em @R=.5em @!R{
\freegate{\bra{j}}
&\gate{H}
&\multigate{1}{\pizz{k}}
&\qw
&\gate{\ket{a}}
\\
&\qw
&\ghost{\pizz{k}}
&\gate{H}
&\gate{\ket{0}}
}
\end{array}
=}\\
&=&
\bra{j}_\bita
H(\bita)
\pizz{k}
\left(\sum_{(a',b')\in Bool^2}
\ket{a',b'}
\bra{a',b'}\right)
H(\bitb)
\ket{a,0}_{\bita\bitb}\\
&=&
\sum_{a',b'}
\bra{j}_\bita
H(\bita)
\ket{a'}_\bita
\ket{b'}_\bitb
\delta^k_{a'\oplus b'}
\bra{a',b'}H(\bitb)
\ket{a, 0}_{\bita\bitb}\\
&=&
\sum_{a',b'}
\frac{(-1)^{ja'}}{\sqrt{2}}
\ket{b'}_\bitb
\delta^k_{a'\oplus b'}
\frac{\delta_a^{a'}}{\sqrt{2}}\\
&=&
\frac{(-1)^{ja}}{2}
\ket{k\oplus a}_\bitb\\
&=&
\sigx^k(\bitb)\sigz^j(\bitb)
\ket{a}_\bitb/2
\;.
\eeqa
\qed
\section{Teleportation}

Throughout this section, $\ket{\psi}$ will denote
an arbitrary one qubit state.

\claim

For any $x, z\in Bool$,

\grayeq{
\beq
2
\begin{array}{c}
\Qcircuit @C=1em @R=.5em @!R{
\freemultigate{1}{\bra{B_{xz}}}
& \gate{\ket{\psi}}
\\
\freeghost{\bra{B_{xz}}}
&\multigate{1}{\ket{B^{xz}}}
\\
&\ghost{\ket{B^{xz}}}
}
\end{array}
=
\begin{array}{c}
\Qcircuit @C=1em @R=.5em @!R{
&
\\
&
\\
&\gate{\ket{\psi}}
}
\end{array}
\;.
\label{eq-tele1}
\eeq}
\proof
Let LHS denote the left
hand side of Eq.(\ref{eq-tele1}).
Then

\beqa
LHS &=&
\bra{B^{00}}_{\bita\bitb}
\sigz^z(\bitb)\sigx^x(\bitb)
\sigx^x(\bitb)\sigz^z(\bitb)
\ket{\psi}_\bita
\ket{B^{00}}_{\bitb\bitc}\\
&=&
\bra{B^{00}}_{\bita\bitb}
\ket{\psi}_\bita
\ket{B^{00}}_{\bitb\bitc}
\;,
\eeqa
so we only need to prove Eq.(\ref{eq-tele1})
for $x=z=0$.

For an arbitrary $a\in Bool$,

\beqa
\bra{B^{00}}_{\bita\bitb}
\ket{a}_\bita
\ket{B^{00}}_{\bitb\bitc}
&=&
(\frac{1}{2})
(\bra{00}_{\bita\bitb} + \bra{11}_{\bita\bitb})
\ket{a}_\bita
(\ket{00}_{\bitb\bitc} + \ket{11}_{\bitb\bitc})\\
&=&
\frac{\bra{a}_{\bitb}}{2}
(\ket{00}_{\bitb\bitc} + \ket{11}_{\bitb\bitc})\\
&=&
\frac{\ket{a}_\bitc}{2}
\;.
\eeqa

Alternatively, note that

\beqa
2
\begin{array}{c}
\Qcircuit @C=1em @R=.5em @!R{
\freemultigate{1}{\bra{B^{00}}}
& \gate{\ket{\psi}}
\\
\freeghost{\bra{B^{00}}}
&\multigate{1}{\ket{B^{00}}}
\\
&\ghost{\ket{B^{00}}}
}
\end{array}
&=&
2
\begin{array}{c}
\Qcircuit @C=1em @R=.5em @!R{
\freegate{\bra{0}}
&\gate{H}
&\dotgate\qwx[1]
&\qw
&\qw
&\gate{\ket{\psi}}
\\
\freegate{\bra{0}}
&\qw
&\timesgate
&\dotgate\qwx[1]
&\gate{H}
&\gate{\ket{0}}
\\
&\qw
&\qw
&\timesgate
&\qw
&\gate{\ket{0}}
}
\end{array}\\
&=&
2
\begin{array}{c}
\Qcircuit @C=1em @R=.5em @!R{
\freegate{\bra{0}}
&\gate{H}
&\dotgate\qwx[1]
&\dotgate\qwx[2]
&\qw
&\gate{\ket{\psi}}
\\
\freegate{\bra{0}}
&\dotgate\qwx[1]
&\timesgate
&\qw
&\gate{H}
&\gate{\ket{0}}
\\
&\timesgate
&\qw
&\timesgate
&\qw
&\gate{\ket{0}}
}
\end{array}\\
&=&
\sqrt{2}
\begin{array}{c}
\Qcircuit @C=1em @R=.5em @!R{
\freegate{\bra{0}}
&\gate{H}
&\dotgate\qwx[2]
&\gate{\ket{\psi}}
\\
&
&
&
\\
&\qw
&\timesgate
&\gate{\ket{0}}
}
\end{array}\\
&=&
\ket{\psi}_\bitc
\;.
\eeqa
\qed

\claim

For any $x, z\in Bool$,

\grayeq{
\beq
2
\begin{array}{c}
\Qcircuit @C=1em @R=.5em @!R{
\freemultigate{1}{\bra{B^{xz}}}
&\qw
& \gate{\ket{\psi}}
\\
\freeghost{\bra{B_{xz}}}
&\qw
&\multigate{1}{\ket{B^{00}}}
\\
&\gate{\sigx^x\sigz^z}
&\ghost{\ket{B^{00}}}
}
\end{array}
=
\begin{array}{c}
\Qcircuit @C=1em @R=.5em @!R{
&
\\
&
\\
&\gate{\ket{\psi}}
}
\end{array}
\;.
\label{eq-tele2}
\eeq}
\proof
Follows immediately from Eq.(\ref{eq-tele1}).
\qed
\section{Dense Coding}

\claim

For any $a,b\in Bool$,

\grayeq{
\beq
\begin{array}{c}
\Qcircuit @C=1em @R=.5em @!R{
&\qw
&\qw
&\dotgate\qwx[2]
&\qw
&\qw
&\qw
&\gate{\ket{a}}
\\
&\qw
&\qw
&\qw
&\dotgate\qwx[1]
&\qw
&\qw
&\gate{\ket{b}}
\\
&\qw
&\timesgate\qwx[1]
&\timesgate
&\dotgate
&\timesgate\qwx[1]
&\qw
&\gate{\ket{0}}
\\
&\gate{H}
&\dotgate
&\qw
&\qw
&\dotgate
&\gate{H}
&\gate{\ket{0}}
}
\end{array}
=
\begin{array}{c}
\Qcircuit @C=1em @R=.5em @!R{
&\gate{\ket{a}}
\\
&\gate{\ket{b}}
\\
&\gate{\ket{a}}
\\
&\gate{\ket{b}}
}
\end{array}
\;.
\label{eq-den-cod}
\eeq}
\proof
Let LHS and RHS denote the left and right
hand sides of Eq.(\ref{eq-den-cod}).

\beq
RHS =
\begin{array}{c}
\Qcircuit @C=1em @R=.5em @!R{
&\gate{\ket{a}}
\\
&\gate{\ket{b}}
\\
\freemultigate{1}{\sum_{(c,d)\in Bool^2}\ket{c,d}\bra{B^{c,d}}}
&\multigate{1}{\ket{B^{a,b}}}
\\
\freeghost{\sum_{(c,d)\in Bool^2}\ket{c,d}\bra{B^{c,d}}}
&\ghost{\ket{B^{a,b}}}
}
\end{array}
\;.
\eeq

\beqa
\begin{array}{c}
\Qcircuit @C=1em @R=.5em @!R{
&\gate{\ket{a}}
\\
&\gate{\ket{b}}
\\
&\multigate{1}{\ket{B^{a,b}}}
\\
&\ghost{\ket{B^{a,b}}}
}
\end{array}
&=&
\ket{a,b}_{\bita\bitb}
\sigx^{a}(\bitc)
\sigz^{b}(\bitc)
\ket{B^{00}}_{\bitc\bitd}\\
&=&
\sigx^{n(\bita)}(\bitc)
\sigz^{n(\bitb)}(\bitc)
\ket{a,b}_{\bita\bitb}
\ket{B^{00}}_{\bitc\bitd}\\
&=&
\begin{array}{c}
\Qcircuit @C=1em @R=.5em @!R{
&\dotgate\qwx[2]
&\qw
&\qw
&\qw
&\gate{\ket{a}}
\\
&\qw
&\dotgate\qwx[1]
&\qw
&\qw
&\gate{\ket{b}}
\\
&\timesgate
&\dotgate
&\timesgate\qwx[1]
&\qw
&\gate{\ket{0}}
\\
&\qw
&\qw
&\dotgate
&\gate{H}
&\gate{\ket{0}}
}
\end{array}
\;.
\eeqa

\beq
\begin{array}{c}
\Qcircuit @C=1em @R=1.5em @!R{
&\qw
\\
&\qw
\\
\freemultigate{1}{\sum_{c,d}\ket{c,d}\bra{B^{c,d}}}
&\qw
\\
\freeghost{\sum_{c,d}\ket{c,d}\bra{B^{c,d}}}
&\qw
}
\end{array}
=
\sum_{c,d}
\begin{array}{c}
\Qcircuit @C=1em @R=.5em @!R{
&\qw
&\qw
&\qw
\\
&\qw
&\qw
&\qw
\\
\freegate{\ket{c}\bra{c}}
&\qw
&\timesgate\qwx[1]
&\qw
\\
\freegate{\ket{d}\bra{d}}
&\gate{H}
&\dotgate
&\qw
}
\end{array}
\;.
\eeq
\qed
\section{Quantum Fourier Transform}

For this section, it is especially
important that
the reader read the Notation section
of QC Paulinesia.
The Notation section explains
what we mean by natural labelling.
Natural labelling will
be used in this section.

Given a vector
$\vec{x}=(x_{\nb-1},\ldots, x_1, x_0 )\in Bool^\nb$,
let
$R\vec{x}=(x_0, x_1, \ldots,x_{\nb-1})$.
Thus $R$ is the matrix that reverses the components
of an $\nb$ dimensional vector.
For example, for $\nb=4$,

\beq
R =
\left(
\begin{array}{cccc}
&&&1\\
\text{\huge 0}&&1& \\
&1&& \\
1&&&\text{\huge 0} \\
\end{array}
\right)
\;.
\eeq
We will also use $R$ to denote
a map from the Hilbert space of $\nb$
qubits to itself such that $R\ket{\vec{x}}=
\ket{R\vec{x}}$ for $\vec{x}\in Bool^\nb$.
We will also use $R$ to denote
the map $R:Z_{0,\nb-1}\rarrow Z_{0,\nb-1}$
such that $R(i)=\nb-1-i$. For example, for
$\nb=4$,
$R$  maps $0\rarrow 3$, $1\rarrow 2$,
$2\rarrow 1$, $3\rarrow 0$.

For any $\bita, \bitb \in Z_{0,\nb-1}$, define

\grayeq{
\beq
\begin{array}{c}
\Qcircuit @C=1em @R=1.5em @!R{
&\dotgate\qwx[1]
&\qw
\\
&\dotgate
&\qw
}
\end{array}
=
V(\bita, \bitb)=
\exp[ i\pi
\frac{n(\bita)n(\bitb)}{2^{|\bita-\bitb|}}
]
=
(-1)^\frac{n(\bita)n(\bitb)}{2^{|\bita-\bitb|}}
\;.
\label{eq-fou-v-def}
\eeq}
Note that normally in QC Paulinesia, we use
$
\begin{array}{c}
\Qcircuit @C=1em @R=1.5em @!R{
&\dotgate\qwx[1]
&\qw
\\
&\dotgate
&\qw
}
\end{array}
=\sigz^{n(\bita)}(\bitb)=
(-1)^{n(\bita)n(\bitb)}
\;,
$
so the definition given by Eq.(\ref{eq-fou-v-def})
applies only to this section.

For any $x\in Z_{0, \ns-1}$, the
{\bf Quantum Fourier Transform}
of $\ket{x}$ is defined by

\grayeq{
\beq
U_{FT}\ket{x}=
\frac{1}{\sqrt{\ns}}
\sum_{y=0}^{\ns-1}
e^{ i\frac{2\pi xy}{\ns} }
\ket{y}
\;.
\eeq}

Henceforth, for simplicity,
we will often assume $\nb=4$.
It will be obvious how to extend our
arguments to other values of $\nb$.

\claim

For any
$\vec{x}=(x_3,x_2,x_1, x_0 )\in Bool^4$,

\grayeq{
\beq
\begin{array}{c}
\Qcircuit @C=1em @R=.5em @!R{
&\freemultigate{3}{U_{FT}}
&\gate{\ket{x_0}}
\\
&\freeghost{U_{FT}}
&\gate{\ket{x_1}}
\\
&\freeghost{U_{FT}}
&\gate{\ket{x_2}}
\\
&\freeghost{U_{FT}}
&\gate{\ket{x_3}}
}
\end{array}
 =
\begin{array}{c}
\Qcircuit @C=1em @R=.5em @!R{
&\qw
&\qw
&\qw
&\qw
&\qw
&\qw
&\dotgate\qwx[3]
&\dotgate\qwx[2]
&\dotgate\qwx[1]
&\gate{H}
&\gate{\ket{x_3}}
\\
&\qw
&\qw
&\qw
&\dotgate\qwx[2]
&\dotgate\qwx[1]
&\gate{H}
&\qw
&\qw
&\dotgate
&\qw
&\gate{\ket{x_2}}
\\
&\qw
&\dotgate\qwx[1]
&\gate{H}
&\qw
&\dotgate
&\qw
&\qw
&\dotgate
&\qw
&\qw
&\gate{\ket{x_1}}
\\
&\gate{H}
&\dotgate
&\qw
&\dotgate
&\qw
&\qw
&\dotgate
&\qw
&\qw
&\qw
&\gate{\ket{x_0}}
}
\end{array}
\;,
\eeq}
\proof

Recall from the Notation section
that $\vec{\nu}=(\nb-1,\ldots,2 , 1, 0)$.
Let
$n=2^{\vec{\nu}}\cdot \vec{n}$ and
$x=2^{\vec{\nu}}\cdot \vec{x}$. Then

\beqa
U_{FT}\ket{\vec{x}}_{\vec{\nu}}&=&
\frac{1}{\sqrt{\ns}}
e^{ \frac{i 2\pi xn}{\ns} }
\sum_{\vec{y}\in Bool^\nb}
\ket{\vec{y}}_{\vec{\nu}}\\
&=&
e^{i \frac{2\pi xn}{\ns} }
H(\vec{\nu})
\ket{0}_{\vec{\nu}}
\;.
\eeqa
Furthermore,

\beqa
\exp[\frac{i 2\pi x n}{16} ]&=&
e^{\left[
\frac{i 2\pi }{16}
(8x_3 + 4x_2 + 2x_1 + x_0  )
(8n(3) + 4n(2) + 2n(1) + n(0))
\right]}\\
&=&
\exp[i 2\pi
\left\{
\begin{array}{l}
\;\;\;n(3)(\frac{x_0}{2})\\
+n(2)(\frac{x_1}{2}+\frac{x_0}{4})\\
+n(1)(\frac{x_2}{2}+\frac{x_1}{4}+\frac{x_0}{8})\\
+n(0)(\frac{x_3}{2}+\frac{x_2}{4}+\frac{x_1}{8}+\frac{x_0}{16})
\end{array}
\right\} ]
\label{eq-fou-omit-two-pi}
\;,
\eeqa
where, in Eq.(\ref{eq-fou-omit-two-pi}),
we omitted all terms
in the argument of the exponential
that
yielded contributions of the form
$e^{i 2\pi (integer)}$.

Note that for any $x\in Bool$
and bit $\bita$,

\beq
(-1)^{xn(\bita)}H(\bita)\ket{0}_\bita=
\sigz^{x}(\bita)H(\bita)\ket{0}_\bita=
H(\bita)\ket{x}_\bita
\;.
\eeq
Thus,

\beqa
\lefteqn{\exp[\frac{i 2\pi x n}{16} ]
H(\vec{\nu})\ket{0}_{\vec{\nu}} =
\left\{
\begin{array}{l}
\exp[i \pi n(3) x_0]H(3)\ket{0}_3\\
\exp[i \pi n(2)(x_1+\frac{x_0}{2})]H(2)\ket{0}_2\\
\exp[i \pi n(1)(x_2+\frac{x_1}{2}+\frac{x_0}{4})]H(1)\ket{0}_1\\
\exp[i \pi n(0)(x_3+\frac{x_2}{2}+\frac{x_1}{4}+\frac{x_0}{8})]H(0)\ket{0}_0
\end{array}
\right.}
\\
&=&
\left\{
\begin{array}{l}
H(3)\ket{x_0}_3\\
\exp[i \pi n(2)(\frac{x_0}{2})]H(2)\ket{x_1}_2\\
\exp[i \pi n(1)(\frac{x_1}{2}+\frac{x_0}{4})]H(1)\ket{x_2}_1\\
\exp[i \pi n(0)(\frac{x_2}{2}+\frac{x_1}{4}+\frac{x_0}{8})]H(0)\ket{x_3}_0
\end{array}
\right.
\\
&=&
\left\{
\begin{array}{l}
H(3)\\
\exp[i \pi n(2)(\frac{n(3)}{2})]H(2)\\
\exp[i \pi n(1)(\frac{n(2)}{2}+\frac{n(3)}{4})]H(1)\\
\exp[i \pi n(0)(\frac{n(1)}{2}+\frac{n(2)}{4}+\frac{n(3)}{8})]H(0)
\end{array}
\right\}R\ket{\vec{x}}
\\
&=&
H(3)
V(3,2)
H(2)
V(3,1)V(2,1)
H(1)
V(3,0)V(2,0)V(1,0)
H(0)
R\ket{\vec{x}}
\;.
\eeqa
\qed

\claim(3-2-1 form equals 1-2-3 form)

\grayeq{
\beqa
U_{FT} &=&
\begin{array}{c}
\Qcircuit @C=1em @R=.5em @!R{
&\qw
&\qw
&\qw
&\qw
&\qw
&\qw
&\dotgate\qwx[3]
&\dotgate\qwx[2]
&\dotgate\qwx[1]
&\gate{H}
&\multigate{3}{R}
\\
&\qw
&\qw
&\qw
&\dotgate\qwx[2]
&\dotgate\qwx[1]
&\gate{H}
&\qw
&\qw
&\dotgate
&\qw
&\ghost{R}
\\
&\qw
&\dotgate\qwx[1]
&\gate{H}
&\qw
&\dotgate
&\qw
&\qw
&\dotgate
&\qw
&\qw
&\ghost{R}
\\
&\gate{H}
&\dotgate
&\qw
&\dotgate
&\qw
&\qw
&\dotgate
&\qw
&\qw
&\qw
&\ghost{R}
}
\end{array}
\label{eq-1-2-3}
\\
&=&
\begin{array}{c}
\Qcircuit @C=1em @R=.5em @!R{
&\qw
&\qw
&\qw
&\dotgate\qwx[3]
&\qw
&\qw
&\dotgate\qwx[2]
&\qw
&\dotgate\qwx[1]
&\gate{H}
&\multigate{3}{R}
\\
&\qw
&\qw
&\dotgate\qwx[2]
&\qw
&\qw
&\dotgate\qwx[1]
&\qw
&\gate{H}
&\dotgate
&\qw
&\ghost{R}
\\
&\qw
&\dotgate\qwx[1]
&\qw
&\qw
&\gate{H}
&\dotgate
&\dotgate
&\qw
&\qw
&\qw
&\ghost{R}
\\
&\gate{H}
&\dotgate
&\dotgate
&\dotgate
&\qw
&\qw
&\qw
&\qw
&\qw
&\qw
&\ghost{R}
}
\end{array}
\;.
\label{eq-3-2-1}
\eeqa}
\proof
Obvious.
\qed

We call ``the 1-2-3 form"
the form of $U_{FT}$ given
by Eq.(\ref{eq-1-2-3}).
We call ``the 3-2-1 form" the
form given by Eq.(\ref{eq-3-2-1}).
The
 numbers 1,2,3
refer to the
number of $V$ operators
between the $H$ operators.

\claim

$U_{FT}$ is a symmetric matrix.
\proof

Let $\dagger=$ Hermitian conjugate,
$*=$ complex conjugate, so $\dagger *=$ transpose.
For any $x,y\in Z_{0,\ns-1}$,

\beq
\bra{y} U_{FT} \ket{x} =
\exp(\frac{i2\pi x y}{\ns}) =
\bra{y} U_{FT}^{\dagger *} \ket{x}
\;.
\eeq
\altproof
\beqa
U_{FT}^{\dagger *} &=&
\begin{array}{c}
\Qcircuit @C=1em @R=.5em @!R{
\freemultigate{3}{R}
&\gate{H}
&\dotgate\qwx[1]
&\dotgate\qwx[2]
&\dotgate\qwx[3]
&\qw
&\qw
&\qw
&\qw
&\qw
&\qw
\\
\freeghost{R}
&\qw
&\dotgate
&\qw
&\qw
&\gate{H}
&\dotgate\qwx[1]
&\dotgate\qwx[2]
&\qw
&\qw
&\qw
\\
\freeghost{R}
&\qw
&\qw
&\dotgate
&\qw
&\qw
&\dotgate
&\qw
&\gate{H}
&\dotgate\qwx[1]
&\qw
\\
\freeghost{R}
&\qw
&\qw
&\qw
&\dotgate
&\qw
&\qw
&\dotgate
&\qw
&\dotgate
&\gate{H}
}
\end{array}\\
&=&
\begin{array}{c}
\Qcircuit @C=1em @R=.5em @!R{
&\qw
&\qw
&\qw
&\dotgate\qwx[3]
&\qw
&\qw
&\dotgate\qwx[2]
&\qw
&\dotgate\qwx[1]
&\gate{H}
&\multigate{3}{R}
\\
&\qw
&\qw
&\dotgate\qwx[2]
&\qw
&\qw
&\dotgate\qwx[1]
&\qw
&\gate{H}
&\dotgate
&\qw
&\ghost{R}
\\
&\qw
&\dotgate\qwx[1]
&\qw
&\qw
&\gate{H}
&\dotgate
&\dotgate
&\qw
&\qw
&\qw
&\ghost{R}
\\
&\gate{H}
&\dotgate
&\dotgate
&\dotgate
&\qw
&\qw
&\qw
&\qw
&\qw
&\qw
&\ghost{R}
}
\end{array}\\
&=& U_{FT}
\;.
\eeqa
\qed

For distinct bits $\bita, \bitb$,
let $V(\bita, \bitb)_{n(\bitb)\rarrow b}$
denote the result of substituting $n(\bitb)$
in
$V(\bita, \bitb)$
by $b\in Bool$.

\claim

For any
$\vec{x}=(x_3,x_2,x_1, x_0 )\in Bool^4$
and $\vec{y}=(y_3,y_2,y_1, y_0 )\in Bool^4$,

\grayeq{\small
\beq
\bra{\vec{y}}U_{FT}\ket{\vec{x}}=
\begin{array}{c}
\Qcircuit @C=1em @R=.5em @!R{
&&&
&\freegate{\bra{y_0}H(0)}
&\gate{\ket{x_{R(0)}}}
\\
&&
&\freegate{\bra{y_1}H(1)}
&\gate{V(1,0)_{n(0)\rarrow y_0}}
&\gate{\ket{x_{R(1)}}}
\\
&
&\freegate{\bra{y_2}H(2)}
&\gate{V(2,1)_{n(1)\rarrow y_1}}
&\gate{V(2,0)_{n(0)\rarrow y_0}}
&\gate{\ket{x_{R(2)}}}
\\
&\freegate{\bra{y_3}H(3)}
&\gate{V(3,2)_{n(2)\rarrow y_2}}
&\gate{V(3,1)_{n(1)\rarrow y_1}}
&\gate{V(3,0)_{n(0)\rarrow y_0}}
&\gate{\ket{x_{R(3)}}}
}
\end{array}
\;.
\eeq}
\proof
Obvious.
\qed

\claim

\grayeq{
\beq
R=
\begin{array}{c}
\Qcircuit @C=1em @R=1em @!R{
&\uarrowgate\qwx[1]
&\uarrowgate\qwx[2]
&\qw
&\uarrowgate\qwx[3]
&\qw
&\qw
&\qw
\\
&\darrowgate
&\qw
&\uarrowgate\qwx[1]
&\qw
&\uarrowgate\qwx[2]
&\qw
&\qw
\\
&\qw
&\darrowgate
&\darrowgate
&\qw
&\qw
&\uarrowgate\qwx[1]
&\qw
\\
&\qw
&\qw
&\qw
&\darrowgate
&\darrowgate
&\darrowgate
&\qw
}
\end{array}
\;.
\label{eq-r-expan}
\eeq}
\proof
Check that the right hand side
of Eq.(\ref{eq-r-expan})
maps
$0\rarrow 3$,
$1\rarrow 2$,
$2\rarrow 1$, and
$3\rarrow 0$.
\qed
\section{References}
The following documents were useful in preparing
this document.

\end{document}